\documentclass[acmtog]{acmart}
\acmSubmissionID{363}

\usepackage[utf8]{inputenc}
\usepackage{booktabs} 
\usepackage[ruled]{algorithm2e} 
\usepackage{tabularx}
\usepackage{multirow}
\usepackage{subcaption}
\usepackage{textcomp}
\usepackage{color}
\usepackage{colortbl}

\begingroup
  \catcode`\_=\active
  \gdef_#1{\ensuremath{\sb{\mathrm{#1}}}}
\endgroup
\mathcode`\_=\string"8000
\catcode`\_=12

\usepackage{xcolor}
\usepackage{textcomp}
\definecolor{gray}{rgb}{0.5,0.5,0.5}
\definecolor{purple}{rgb}{0.7,0.3,0.7}
\definecolor{blue}{rgb}{0,0,1}
\definecolor{darkblue}{rgb}{0,0,0.6}
\definecolor{orange}{rgb}{1,.5,0} 
\definecolor{red}{rgb}{1,0,0} 

\newcommand{\changed}[1]{#1}

\newcommand{\fig}[1]{Figure~\ref{#1}}

\newcommand{\Mat}[1]    {{\ensuremath{\mathbf{\uppercase{#1}}}}} 

\newcommand{\normone}[1]{\left\lVert#1\right\rVert_1}
\newcommand{\normtwo}[1]{\left\lVert#1\right\rVert_2}

\newcommand{\mask}{\Mat{M}}
\newcommand{\solidangle}{\Mat{\Omega}}

\newcommand{\image}{\Mat{I}}
\newcommand{\light}{\Mat{L}}
\newcommand{\sourceimage}{\Mat{I}_{\textrm{s}}}
\newcommand{\targetlight}{\Mat{L}_{\textrm{t}}}
\newcommand{\targetimage}{\Mat{I}_{\textrm{t}}}
\newcommand{\sourcelight}{\Mat{L}_{\textrm{s}}}

\newcommand{\imagepred}{\hat{\Mat{I}}}
\newcommand{\lightpred}{\hat{\Mat{L}}}
\newcommand{\sourceimagepred}{\hat{\Mat{I}}_{\textrm{s}}}
\newcommand{\targetimagepred}{\hat{\Mat{I}}_{\textrm{t}}}
\newcommand{\sourcelightpred}{\hat{\Mat{L}}_{\textrm{s}}}

\newcommand{\loss}{\mathcal{L}}
\newcommand{\imageloss}{\mathcal{L}_{\textrm{I}}}
\newcommand{\lightloss}{\mathcal{L}_{\textrm{L}}}

\newcommand{\relight}{\Mat{\Phi}}

\def\etc{\textit{et al.}}

\newcommand{\selfweight}{\lambda_{\mathit{self}}}
\newcommand{\lightweight}{\lambda_{\mathit{light}}}

\setcitestyle{square,nosort}

\SetAlFnt{\small}
\SetAlCapFnt{\small}
\SetAlCapNameFnt{\small}
\SetAlCapHSkip{0pt}
\IncMargin{-\parindent}

\setcopyright{acmlicensed}
\acmJournal{TOG}
\acmYear{2019}\acmVolume{38}\acmNumber{4}\acmArticle{79}\acmMonth{7} \acmDOI{10.1145/3306346.3323008}

\newcommand{\ignore}[1]{}

\begin{document}
\title{Single Image Portrait Relighting}


\author{Tiancheng Sun}
\email{tis037@cs.ucsd.edu}
\affiliation{\institution{University of California, San Diego}}
\author{Jonathan T. Barron}
\email{barron@google.com}
\author{Yun-Ta Tsai}
\email{yuntatsai@google.com}
\affiliation{\institution{Google Research}}
\author{Zexiang Xu}
\email{zexiangxu@cs.ucsd.edu}
\affiliation{\institution{University of California, San Diego}}
\author{Xueming Yu}
\email{xuemingyu@google.com}
\author{Graham Fyffe}
\email{fyffe@google.com}
\author{Christoph Rhemann}
\email{crhemann@google.com}
\author{Jay Busch}
\email{jbusch@google.com}
\author{Paul Debevec}
\email{debevec@google.com}
\affiliation{\institution{Google}}
\author{Ravi Ramamoorthi}
\email{ravir@cs.ucsd.edu}
\affiliation{\institution{University of California, San Diego}}

\begin{abstract}
Lighting plays a central role in conveying the essence and depth of the subject in a portrait photograph.
Professional photographers will carefully control the lighting in their studio to manipulate the appearance of their subject, while consumer photographers are usually constrained to the illumination of their environment.
Though prior works have explored techniques for relighting an image, their utility is usually limited due to requirements of specialized hardware, multiple images of the subject under controlled or known illuminations, or accurate models of geometry and reflectance.
To this end, we present a system for \emph{portrait relighting}: a neural network that takes as input a \emph{single} RGB image of a portrait taken with a standard cellphone camera in an unconstrained environment, and from that image produces a relit image of that subject as though it were illuminated according to any provided environment map.
Our method is trained on a small database of 18 individuals captured under different directional light sources in a controlled light stage setup consisting of a densely sampled sphere of lights.
Our proposed technique produces quantitatively superior results on our dataset's validation set compared to prior works, and produces convincing qualitative relighting results on a dataset of hundreds of real-world cellphone portraits. Because our technique can produce a $640 \times 640$ image in only $160$ milliseconds, it may enable interactive user-facing photographic applications in the future.
\end{abstract}

%
%


%
%

\newcommand{\teaserwidth}{1.7in}
\begin{teaserfigure}
    \begin{subfigure}[b]{.25\textwidth}
    \centering
    \includegraphics[width=\teaserwidth]{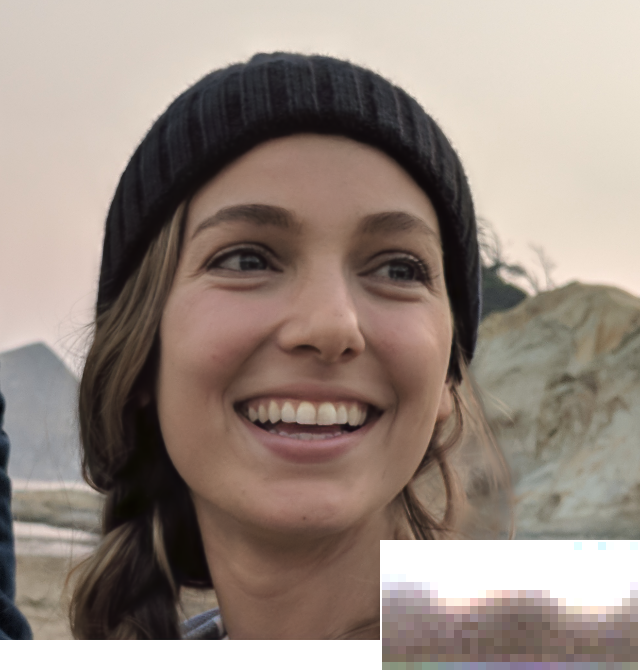}
    \caption{Input image and estimated lighting}\label{subfig:teaser_input}
    \end{subfigure}
    \begin{subfigure}[b]{.75\textwidth}
    \centering
    \includegraphics[width=\teaserwidth]{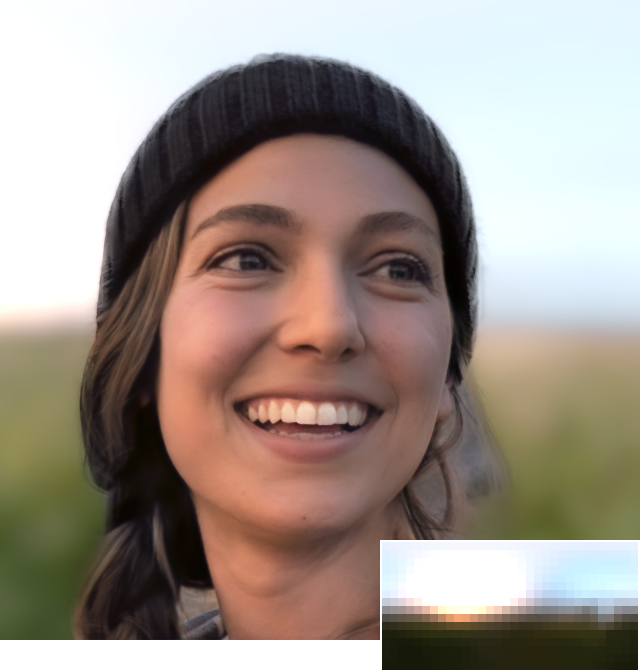}
    \includegraphics[width=\teaserwidth]{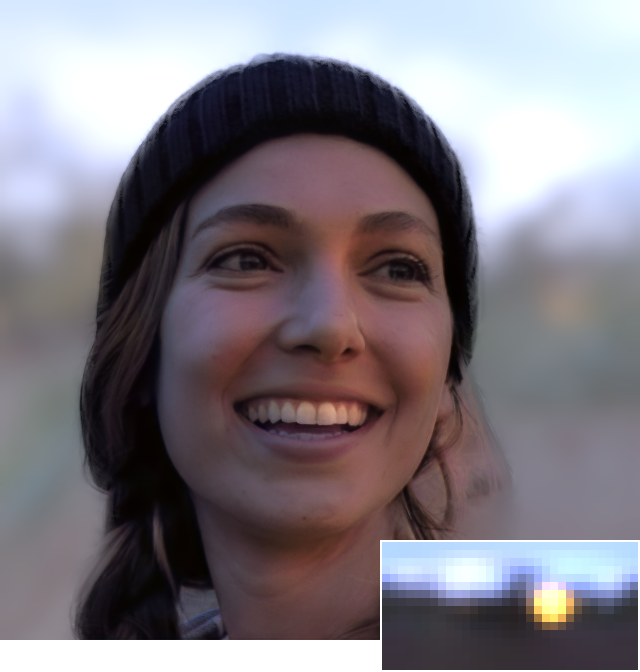}
    \includegraphics[width=\teaserwidth]{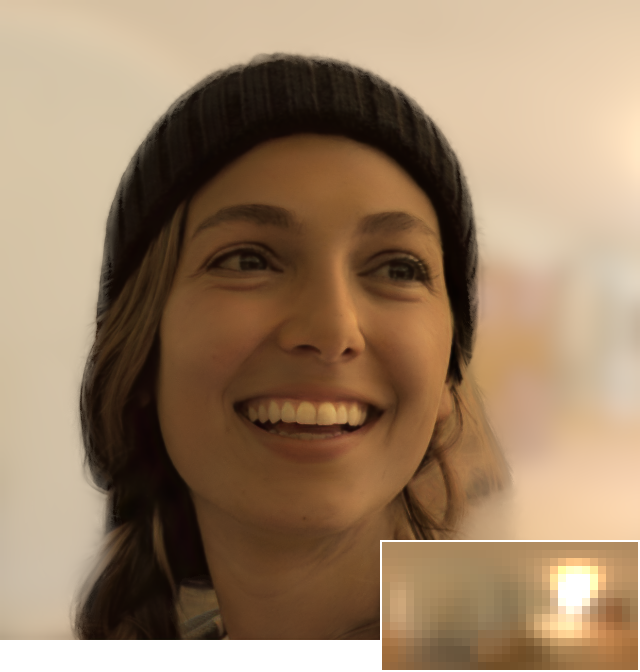}
    \caption{Rendered images from our method under three novel illuminations}\label{subfig:teaser_output}
    \end{subfigure}
    \vspace*{-.2in}
    \caption{Given only a single input image taken with a standard cellphone camera of a portrait (\subref{subfig:teaser_input}), our model is able to quickly ($160$ ms.) generate new images of our human subject as though they are illuminated under new, previously-unseen lighting environments (\subref{subfig:teaser_output}).
    }
    \label{fig:teaser}
\end{teaserfigure}

\maketitle

\section{Introduction}

The rise of mobile computing has led to tremendous growth in the popularity of consumer digital photography, and one of the most popular and ubiquitous kinds of photos taken is the portrait: an image of a human subject's face or upper body.
\changed{Portrait photography follows in the tradition of portrait painting, where since the renaissance artists have recognized how lighting can capture the depth and essence of the subject on a 2D canvas~\cite{schutze2015caravaggio}.
These ideas largely influenced professional portrait lighting techniques~\cite{schriever1909complete}. In the 18th and 19th centuries, portraits were often taken by professional photographers, who carefully considered and controlled the lighting of the scene in addition to the pose and appearance of their subject. 
In the modern age of ``selfies'' and candid photography, it is difficult or impossible for consumer photographers to control the lighting of their subject --- one would likely not consider interrupting an engagement proposal or a piano recital to set up studio lighting.
Moreover, professional-quality lighting often requires expertise and special equipment such as flashes, deflectors and diffusion panels. 
These techniques are out of reach for casual mobile phone photographers, whose only available light is usually the flash built into standard consumer cameras, which tends to produce garishly-lit results~\cite{Petschnigg2004}.}
These casual photographers would benefit greatly from the ability to {\em relight} their portrait photos, and change the lighting of their image in a post-processing setting to that of some other environment.

In this paper, we present a model that requires as input only a {\em single RGB image} of a portrait of a single human subject casually captured with a cellphone camera in natural, unconstrained lighting (\fig{subfig:teaser_input}).
Our model consists of a single deep neural network that has been trained to take such an image as input and produces as output a relit version of the portrait under an arbitrary user-specified environment map (\fig{subfig:teaser_output}), in addition to predicting the environment map corresponding to the input image (\fig{subfig:teaser_input}).

\begin{figure}
\centering

\begin{subfigure}[b]{\linewidth}
\centering
\caption{Complete Relighting}\label{subfig:usecase1}
\includegraphics[width=\linewidth]{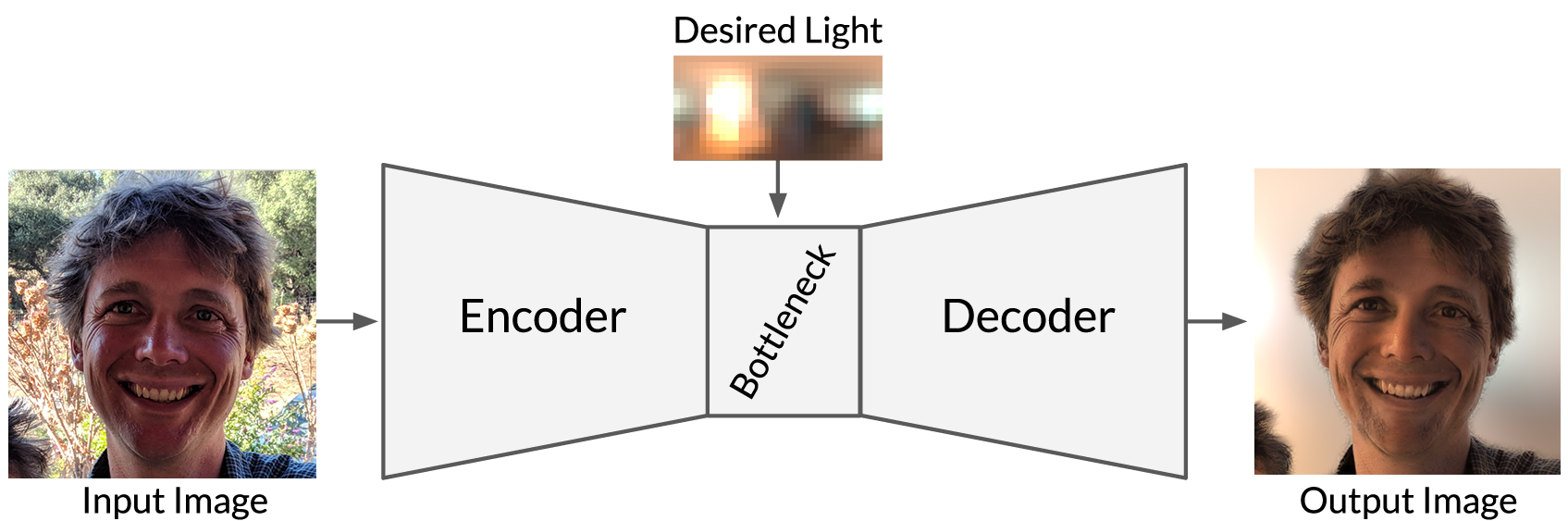}
\end{subfigure}
\begin{subfigure}[b]{\linewidth}
\centering
\caption{Illumination Retargeting}\label{subfig:usecase2}
\includegraphics[width=\linewidth]{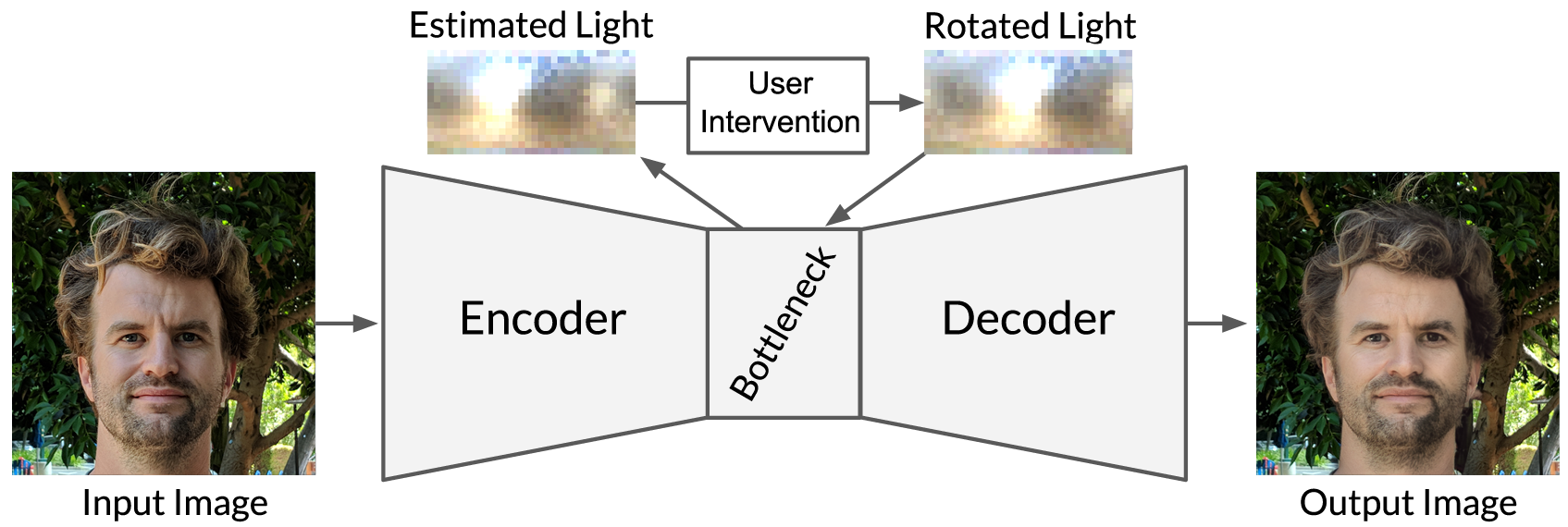}
\end{subfigure}
\vspace*{-.1in}
\caption{
Our relighting system is an encoder-decoder neural network that takes as input a single input image and a target illumination (injected into the bottleneck of the network), and produces as output a relit image (\subref{subfig:usecase1}).
The encoder also predicts the illumination of the input image, thereby allowing an input image's illumination to be recovered during encoding, modified within the bottleneck of the network, and then decoded to produce a relit image corresponding to, say, a rotation of the original illumination (\subref{subfig:usecase2}).
}
\label{fig:system_architecture}
\vspace*{-.1in}
\end{figure}

Unlike traditional~\cite{BarronTPAMI2015} and learning-based~\cite{sfsnetSengupta18} scene inference or face reconstruction algorithms, our model does not have any explicit inverse rendering step for estimating geometry and reflectance.
Such approaches are necessarily limited to what is expressible by their estimated model, such as Lambertian reflectance and spherical harmonic illumination.
In contrast, we train a single neural network (\fig{fig:system_architecture}) to directly produce relit images from an input image and a ``target'' illumination --- any representation of geometry and reflectance that our model may recover is fully learned and is represented solely in terms of internal network activations.
By training our network to directly predict a final image and by not imposing any physical constraints on the implicit representation of our underlying scene (other than the constraints imposed by neural networks which, given enough capacity, are known to be universal function approximators), our model is capable of capturing most of the subtleties of human facial appearance, such as shading, scattering, specularities, and shadowing.

One could imagine our proposed model as the core component of a user-facing system designed to enable post-capture illumination manipulation on a mobile phone. The user could relight their portraits using various canonical or user-specified environments (\fig{subfig:usecase1}), similar to the popular ``filters'' used by social media platforms like Instagram. Or the user could relight their image with the environment map corresponding to the input image produced by our model, by manually rotating or retargeting the environment map to, for example, turn a side-lit scene into a front-lit scene (\fig{subfig:usecase2}).
In this way, our model is superficially similar to the ``Portrait Modes'' already in use by Apple's iPhone~\cite{ApplePortraitMode} and Google's Pixel phone~\cite{Barron2015A, Wadhwa2018} which use learning and computer vision techniques to enable post-capture manipulation of a camera's physical depth-of-field.
Additionally, the iPhone X features a ``Portrait Lighting'' mode that adjusts the brightness and contrast of a human subject, and darkens the background.
Critically, this feature does not \emph{relight} a scene --- it cannot remove or change the illumination of an input image; it can only modify the effects of the existing illumination.

Our method is trained and validated using only real light stage data for just 22 individuals captured from 7 different viewpoints and under 304 lighting directions (18 individuals are used for training, and the remaining 4 are used to validation).
By training on real data, we can preserve the subtleties of non-Lambertian reflectance as it relates to the intricate geometry of human faces.
This dataset is used to train a novel neural network architecture for portrait relighting, which consists of an encoder-decoder structure that has been augmented to allow for the input image's illumination to be predicted from the bottleneck of the network, and allows for the target illumination that is to be used for relighting to be injected into the bottleneck of the network.
Finally, we demonstrate convincing ``in the wild'' results on portrait relighting from a single input image, on a dataset of hundreds of casually captured cellphone portraits.  

\section{Related Work}

Portrait relighting can be seen as a special case of image relighting, and also relates to intrinsic images and shape from shading, as well as monocular face and scene reconstruction methods, style transfer, photographic post-processing, and deep learning.  

{\bf Image-Based Relighting: } The linearity of light transport allows one to take many images of a subject from the same viewpoint under different known illuminations, and relight the subject simply by weighting and linearly combining those images~\cite{debevec2000acquiring}.  Though effective, these techniques require hundreds of images, a near-motionless subject, and precise control over illumination. Therefore, they do not provide a solution when given a single RGB image of an unknown subject in an unconstrained environment. However, we show that, just as a light stage scan of a person can be used to relight that person under any environment, the data collected from such a light stage can be used for training and evaluating our own portrait relighting algorithm.

{\bf Intrinsic Images and Shape from Shading: } ``Intrinsic image'' algorithms, such as the Retinex method~\cite{Land71lightnessand}, abstract away geometry and attribute image content to either shading or reflectance. But these solutions do not enable general purpose relighting, as they do not produce the scene geometry required for rendering novel illuminations.  Shape-from-shading solutions assume that materials and illumination are known or fixed and attempt to recover only geometry~\cite{Horn1970}. But even this constrained problem formulation suffers from a variety of ambiguities~\cite{Belhumeur1999}. We compare to a recent work~\cite{BarronTPAMI2015}, which uses priors to estimate shape, illumination, and reflectance from a single image, and we demonstrate that our algorithm has superior relighting performance.

{\bf Monocular Scene \changed{Understanding}: } Monocular 3D estimation techniques~\cite{Hoiem2007, Saxena2009} are capable of recovering scene geometry from a single image, and though such geometry allows for lights to be synthetically \emph{added} to a scene through straightforward rendering techniques, it does not provide a means by which lights can be \emph{removed}, and therefore does not enable \emph{re}lighting. 
Similarly, monocular illumination estimation techniques~\cite{lalonde-iccv-09,gardner-sigasia-17,holdgeoffroy-cvpr-17} can recover the environment map that illuminated a single image, but this only enables the lighting of 3D geometry that is added to the scene and does not solve the problem of relighting existing scene content.

{\bf Reflectance and Style Transfer:}
Peers \etc~\shortcite{Peers:2007:PFP} showed that a flat-lit facial portrait could be re-lit according to a target environment by multiplying it by an aligned image of a similar reference face seen in the target lighting environment, divided by the appearance of that reference face seen in flat lighting.
In this case, high-frequency lighting changes in the skin and hair are transferred from the reference subject, affecting the appearance of hair and specular reflections. The technique requires a database of reference subjects (chosen manually to approximate the desired subject) to approximate the facial portrait. Shih \etc~\shortcite{Shih:2014:STH} uses a multiscale technique to transfer the local image statistics of a reference facial portrait onto a new one, matching properties such as local contrast and the overall lighting direction. This technique can produce compelling results, but is limited in how much it can change the lighting, and can require manual touch-ups.  Nonetheless, both of these techniques establish that believable relighting can be achieved through image processing operations based on reference images of other subjects, an idea which is at the core of our technique as well.

{\bf Face Reconstruction and Relighting: } \changed{Using a morphable model of faces allows for relighting} under certain circumstances~\cite{blanz1999morphable}. But relying on a highly specialized model of faces results in poor performance when presented with non-generic faces or expressions, or with the non-face content that is common in portraits such as glasses, shoulders, or hair. The method of Shu \etc~\shortcite{shu2018portrait} builds upon these morphable models to specifically target single-image portrait lighting transfer with compelling results, though the proposed system does not allow for general purpose relighting with any environment map, and shares the aforementioned limitations of a basic morphable model-based approach.

{\bf Photographic Post-Processing: } 
Relighting can be thought of as a natural extension of the photometric manipulation already performed by photographic image processing pipelines. Modern digital cameras carefully consider and modify the brightness, exposure, and tone of their output images, all of which can be thought of as the coarse manipulation of scene lighting~\cite{Hasinoff2016}.
Local tone-mapping, which aims to decrease the dynamic range of an input image while preserving its local contrast, can also be thought of as a coarse approximation to relighting, and has been explored through classic algorithms~\cite{FFLS-2008,Paris2011} and learning-based approaches~\cite{GharbiSIGGRAPH2017}.

{\bf Deep Learning: }
The revival of end-to-end trained neural networks in the ongoing ``deep learning'' boom, propelled by compelling results in object recognition~\cite{Krizhevsky2012} and enabled by the availability of differentiable programming frameworks such as Tensorflow~\cite{Tensorflow}, has changed the research landscape across computer science.
In particular, convolutional neural networks~\cite{LeCun1989} have become critical tools in a variety of works related to relighting.
The network proposed by Xu \etc~\shortcite{xu2018deep} is able to learn a relighting function that can reproduce complicated illumination effects, but requires five images captured under predefined directional lights as input.
Sengupta \etc~\shortcite{sfsnetSengupta18} leverages 3D morphable models as a source of synthetic training data to train a network to regress from a face image to a Lambertian intrinsic decomposition and spherical harmonic illumination, which can then be used for relighting. However, the modeling assumptions imposed by this technique limit the quality of its output renderings, as skin is highly non-Lambertian and real-world environment illuminations are poorly approximated by low-order spherical harmonic illumination.
There also exist neural networks~\cite{calian2018face2light, gardner-sigasia-17,holdgeoffroy-cvpr-17} that regress from an image (of a face, of an indoor environment, or of an outdoor environment respectively) to the environment that the image was illuminated by, but these approaches do not provide a solution for relighting.
Li \etc~\shortcite{li2018learning} learns to regress from a single image of an object to a shape and spatially-varying BRDF that can then be used to relight that object, but requires a known and constrained illumination in the form of a front-facing flash, thereby preventing user manipulation of any existing natural illumination of the scene, and limiting the system's utility to situations in which the camera flash is the dominant light source and is not socially-disruptive.
\changed{In this paper, we use a neural network to directly learn a function for relighting images in an end-to-end fashion, without explicitly modeling the geometry or the reflectance of the human face.}

\section{Learning Relighting for Portraits}\label{sec:model}

Given a single high-resolution portrait image taken under a natural environment, we want to change only the illumination of the image to another specified lighting environment while keeping the subject of the portrait image the same.
We assume that the portrait is taken roughly from the front-facing direction, within \textpm$20$ degrees of deviation from the center, and that the lights are distant from the subject.
In addition to a relit image, we want our model to predict the illumination of the input portrait image. Formally, given the ``source'' portrait image $\sourceimage$ and the ``target'' lighting $\targetlight$, we want to learn a function $\relight(\cdot)$ that predicts the lighting of the source portrait image  $\sourcelightpred$ and a target portrait image $\targetimagepred$ lit by the target lighting:
\begin{equation}\label{equ:relight}
    \targetimagepred, \sourcelightpred = \relight(\sourceimage, \targetlight).
\end{equation}
We want $\relight(\cdot)$ to generalize to portraits of a wide range of people with different skin tones, facial expressions, and viewing directions, all taken under a wide variety of different far-field illuminations.

Previous works~\cite{BarronTPAMI2015,sfsnetSengupta18,kanamori2018relighting} usually address this problem by first performing an intrinsic decomposition on the image, and then using that decomposition to render a new image according to the target lighting. This approach places implicit constraints on the resulting rendering, often in the form of Lambertian reflectance and spherical harmonic illumination. In this paper, we adopt an end-to-end approach: we use a neural network to model the relighting function $\relight(\cdot)$ that directly predicts the portrait under the target lighting. By using learning instead of an explicit representation of a reconstructed scene, our approach is capable of modeling the broader range of physical phenomena present in human skin and faces including interreflection, scattering, shading, and specularities.

In Section~\ref{sec:model-data}, we describe our pipeline for portrait collection and dataset synthesis. The resulting portrait input/output pairs are then used as training data by our Portrait-Relighting Network, which is described in Section~\ref{sec:model-net}. Our loss function and our training procedure are described in Section~\ref{sec:model-train}.

\begin{figure}
  \centering
  \begin{subfigure}[t]{0.46\linewidth}
    \centering
    \includegraphics[width=\linewidth]{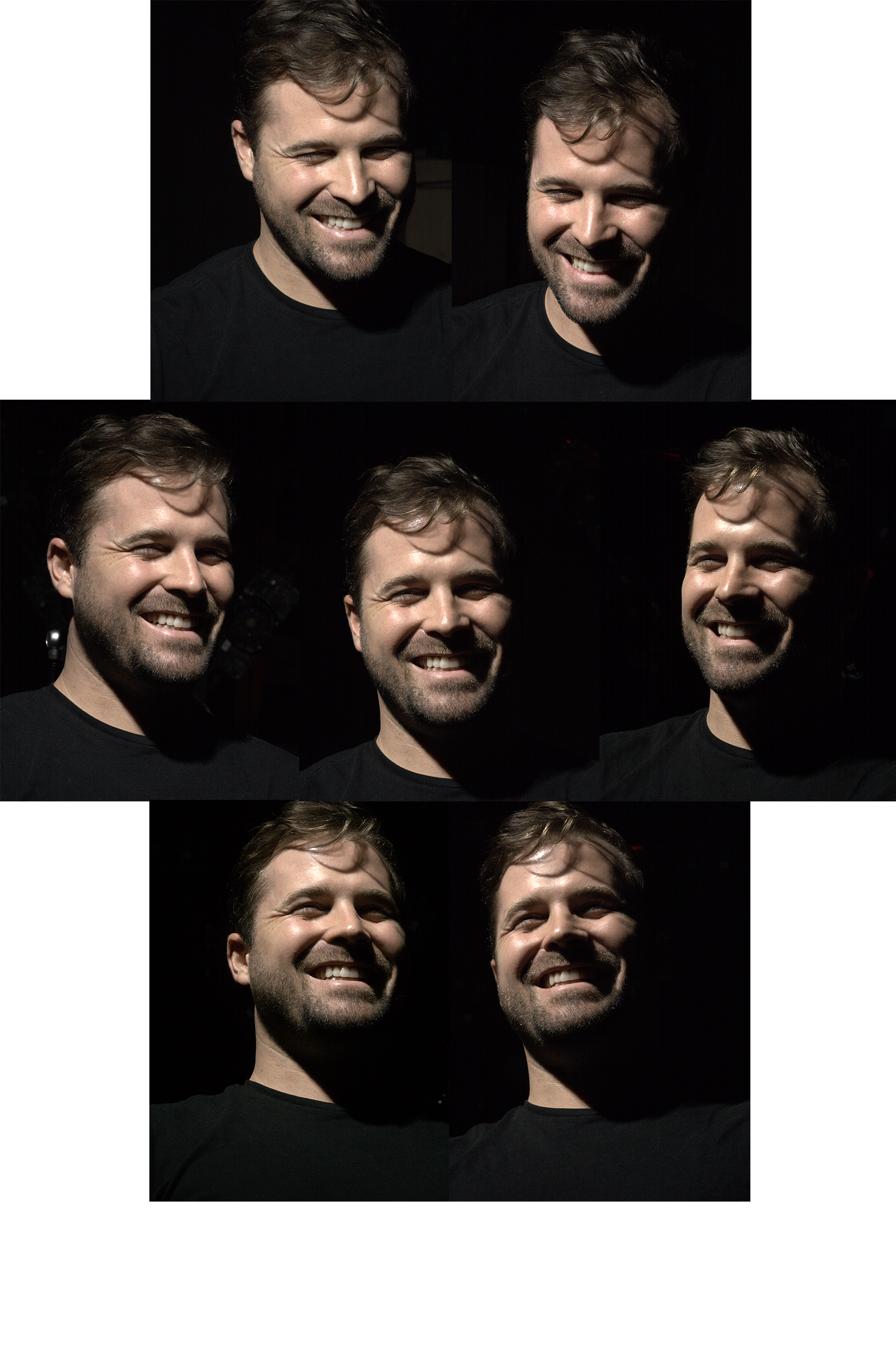}
    \caption{OLAT images (7 cameras).}
    \label{subfig:camera_rig}
  \end{subfigure}
  \quad
  \begin{subfigure}[t]{0.46\linewidth}
    \centering
    \includegraphics[width=\linewidth]{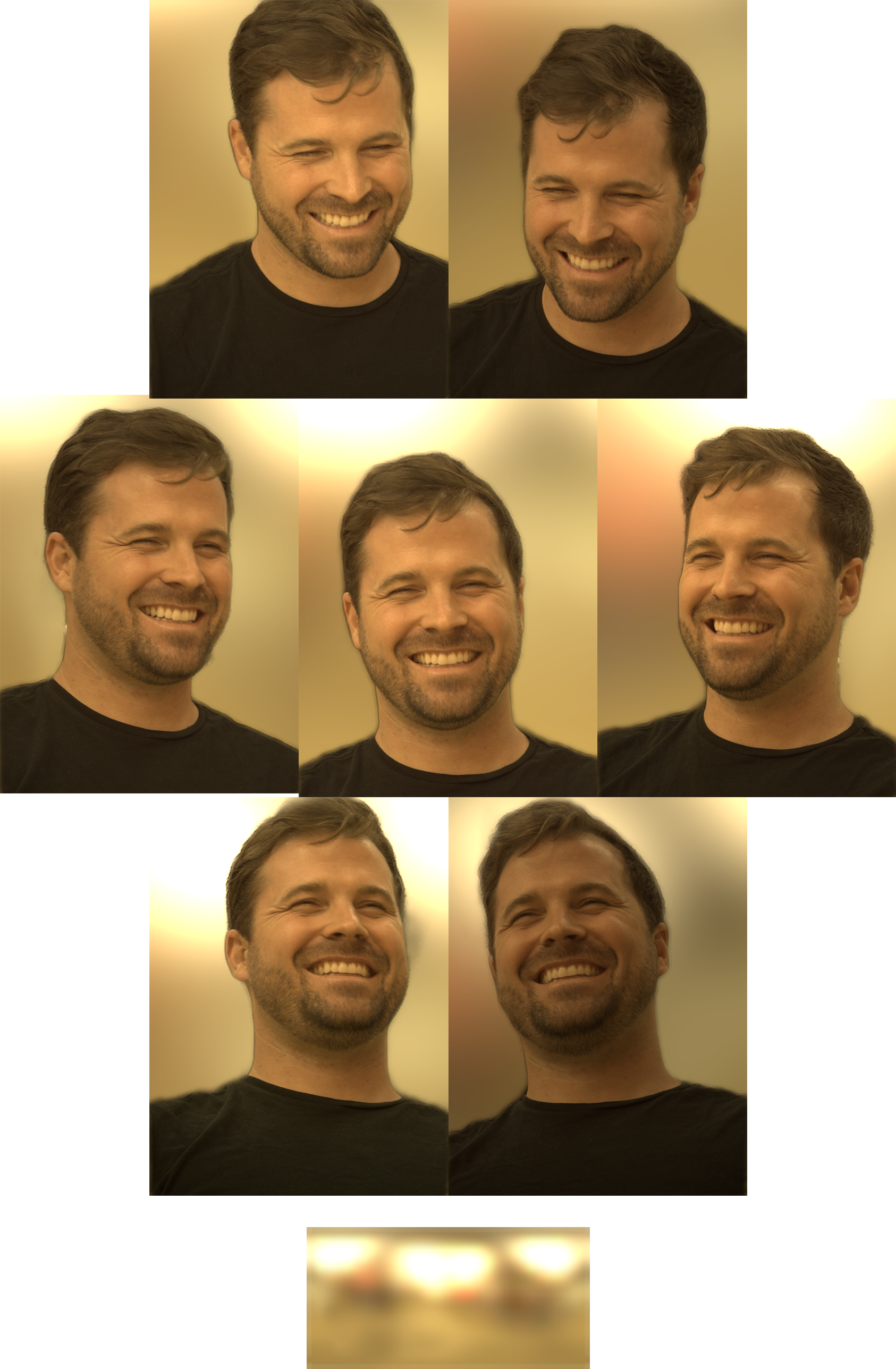}
    \caption{Ground-truth renderings.}
    \label{subfig:relit}
  \end{subfigure}
  \caption{A set of example images from the hexagon shaped multi-camera rig installed on our light stage for one light in an OLAT imageset (\subref{subfig:camera_rig}). The center camera is 1.7 meters away from the subject, and each vertex camera deviates from the center camera by around 20 degrees. This allows us to cover the most common views of a human face in a portrait or selfie. Then, the images can be linearly weighted and combined to obtain images lit by environment illuminations, to generate ground truth for training (\subref{subfig:relit}).}
  \vspace*{-.1in}
  \label{fig:multi-camera_rig}
\end{figure}

\begin{figure*}
    \centering
    \includegraphics[width=\linewidth]{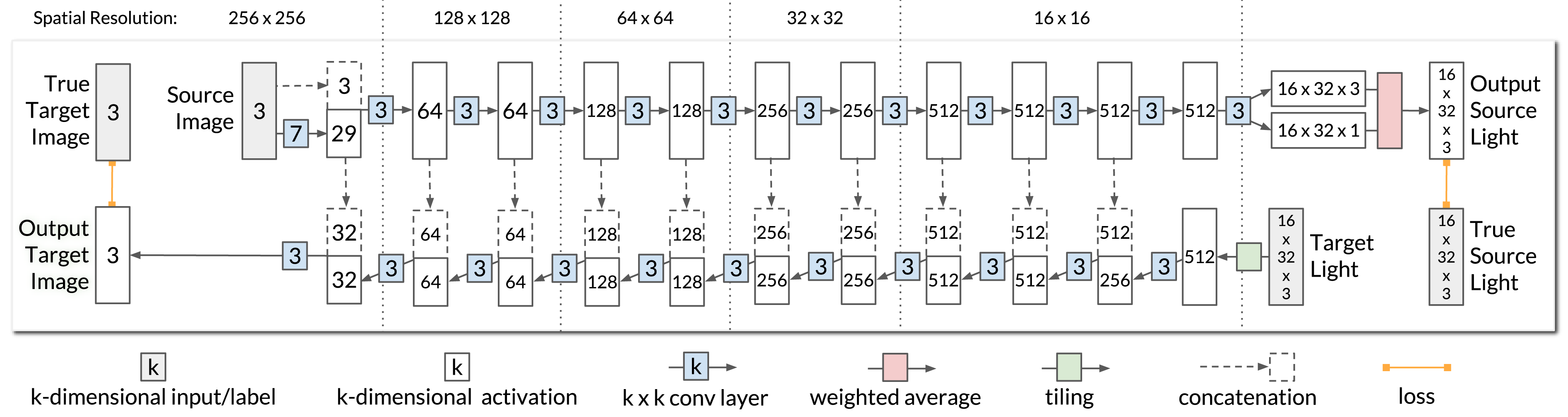}
    \vspace*{-.2in}
    \caption{The architecture of our neural network.
    The source input image is passed through a series of conv layers that gradually decrease spatial resolution while increasing the number of channels. After encoding, a confidence-weighted average predicts the illumination that the source image was lit by. The target light is then injected as input into the bottleneck of the network, and this target light along with the encoding of the source image is decoded (with skip connections) into the output target image. Losses are imposed to minimize the differences between the true and output target images, and between the true and output source lights.
    When evaluating our ``self-supervision'' loss, this architecture is modified such that the output source light with a certain rotation is used as the target light when decoding.}
    \vspace*{-.1in}
    \label{fig:architecture}
\end{figure*}
\subsection{Data}\label{sec:model-data}

{\bf Portrait Imagery:}
In contrast to most supervised learning tasks in computer vision, collecting even a small amount of data for a relighting task is onerous:
the subject must be stationary, and must be repeatedly imaged from the same viewpoint with different known illuminations.
For this reason, previous works~\cite{sfsnetSengupta18,li2018learning} have relied heavily on synthetic, often Lambertian, images for use as training data.
However, there is a significant gap between real and synthetic faces, particularly in terms of skin reflectance~\cite{Donner2008} and fine geometric detail~\cite{Nagano2015}.
And though morphable models~\cite{blanz1999morphable} can be used for synthesizing images of faces, they do not provide a means for synthesizing realistic hair, torsos, or accessories.
The efficacy of modern deep learning techniques means that, though neural networks are capable of learning a complicated mapping from input to output, they will also absorb the biases and inaccuracies of the data with which they have been trained.
Because we want to produce more realistic output than the Lambertian synthetic data commonly used for training, we choose to use only real-world portraits when learning $\relight(\cdot)$.

We use a light stage setup similar to Wenger \etc~\shortcite{Wenger:2005:PRR} to capture the real-world portrait images that will be used to synthesize our training data.
The light stage consists of 304 LED light sources tessellating a sphere surrounding the subject, and seven machine vision cameras positioned in front of the subject, where all lights and cameras are programmable and synchronized. A human subject is seated in the center of the light stage and is asked to remain stationary. We capture a series of images of the subject in which each of the 304 LED light sources is turned on to a white color while all other lights are off. We refer to this set of images produced by each camera as a ``one-light-at-a-time'' (OLAT) imageset. The acquisition process lasts about 6 seconds, during which all but the most disciplined of subjects have likely moved slightly.
For this reason, during each capture we take an additional image every 11th frame in which all LEDs are on.
These ``tracking frames'' are used as input to an optical flow algorithm~\cite{Wenger:2005:PRR}, and the resulting flow fields from these frames are used to advect the OLAT frames in the imageset (assuming linear motion between tracking frames) to remove all but the smallest movements made by the subject.

A 7-camera rig (\fig{fig:multi-camera_rig}(a)) in a hexagonal shape is installed on the light stage, and covers a viewing angle for a human face of \textpm20 degrees with respect to the frontal view. We capture 22 subjects, each with 3 to 5 different facial expressions (neutral, smile, laugh, frown, and raised eyebrows). We use 18 of the subjects for training and the rest for validation.
The demographics of our subjects were largely determined by circumstance, and as such they are somewhat skewed in terms of skin color and gender. We therefore manually constructed our training and validation set in an effort to prevent overfitting and to avoid under-representing minorities in the validation set.
Our training set contains 7 male Caucasians, 7 male Asians, 2 female Caucasians, 1 female Asian, and 1 female of African descent. 
Our validation set contains 1 male Caucasian, 2 male Asians, and 1 female of African descent.
The combination of subjects and facial expressions results in 98 captures, with 81 used for training.

{\bf Environment Lighting:}
To render our OLAT imagesets under natural illumination conditions for training and testing purposes (\fig{fig:multi-camera_rig}(b)), we require many natural illuminations. For this, we use the Laval Indoor HDR Dataset~\cite{gardner-sigasia-17} which consists of 2144 high dynamic range indoor environment maps, and we further collect 950 high dynamic range outdoor environment maps taken from a variety of sources: Eisklotz, HDRI Haven, HDRI Skies, HDRLabs, HDRMAPS, NoEmotion HDRs, and Openfootage. In all, we have 3094 different environment illuminations, of which a random selection of 2560 are used for training and the remaining 534 are used for validation.

{\bf Data Processing:}
With our OLAT imagesets and our environment illuminations, we can now generate paired portrait images which differ only in terms of illumination.
The resolution of our OLAT images is $2400 \times 1800$.
For each image, we first use the human segmentation algorithm~\cite{Wadhwa2018} to mask out the background.
Then we take a random crop from the OLAT imageset, with a position chosen uniformly at random and with the crop size chosen uniformly from $[512, 1024]$ at random.
We then randomly select two environment illuminations (represented in latitude-longitude format), and to each we apply a random rotation in $[0^\circ, 360^\circ]$ in its longitude (but not in its latitude).
Each environment illumination is then resized to $128\times 256$, at which point it is projected onto the basis defined by the LEDs of the light stage.
By taking a weighted combination of the images in the OLAT imageset according to these projections onto this LED basis, we are able to render the cropped OLAT imageset under our two lighting conditions.
After rendering, these illuminations are then mapped back to the latitude-longitude format of size $16 \times 32$.
This is done by modelling each LED as a Gaussian light (where mean position $\mu$ is each LED's position in camera coordinates and standard deviation $\sigma=8^\circ$) and then projecting onto the basis defined by the PDF of these Gaussians.
Finally, the rendered portrait pair is downsampled to $256 \times 256$, and each image (and its corresponding environment illumination) is scaled such that the maximum pixel intensity in each image is $1$.
All processing is done in the linear domain, without gamma correction.
In total, we generate 226,800 portrait pairs as training data.
Note that there is no overlap between training or validation subjects or environment illuminations --- training illuminations are only ever paired with training OLAT imagesets, and validation illuminations are only ever paired with validation OLAT imagesets.

\subsection{Network Architecture}\label{sec:model-net}

The structure of our Portrait-Relighting Network (PR-Net) is shown in~\fig{fig:system_architecture} and \fig{fig:architecture}, and superficially resembles the structure of the popular U-Net~\cite{ronneberger2015u}: an encoder-decoder structure with skip connections.
In the encoder, the source portrait image $\sourceimage$ is processed by a series of conv layers (with strides of 1 or 2) that gradually decrease spatial resolution and increase the number of channels.
The final encoded activations at the bottleneck of the network are then used to predict the illumination corresponding to the source image $\sourcelightpred$.
This illumination prediction is performed using the ``confidence learning'' approach~\cite{hu2017fc4}, which has previously been used for the related problem of color constancy.
This approach is used because, due to the finite spatial support of the conv layers in our network, the encoded network activations only reflect a limited region of the whole image --- they can only ``see'' what lies within their receptive field.
If the network only observes the part of the image corresponding to the left cheek of subject, it's unreasonable to ask it to predict the incident illumination coming from the right.
However, if we designed our network architecture such that the left side of the illumination were to be predicted solely from the left side of the image, we would be depriving our network of the ability to reason about the entire image, which is valuable given the global nature of illumination.
Therefore in our lighting prediction, 
for each location in our encoding bottleneck (each of which corresponds to some limited receptive field) we have our network predict a complete RGB environment illumination as well as a confidence \changed{map} associated with that location in the bottleneck.
We then take a weighted average of these predictions according to their corresponding confidences to get our final predicted lighting $\sourcelightpred$. \changed{Note that in order to keep our network fully convolutional, the spatial resolution of the environment illumination is packed into the channel dimension.}

In the decoder of our network, we feed the target lighting $\targetlight$ as input into the network and encode it using conv layers before concatenating it with the representation of the source image produced by the encoder.
This concatenated encoding is then passed through a series of transposed conv layers (with strides of 1 or 2), and is gradually upsampled back to a 3-channel image with the spatial resolution of the input source image, which is our output target portrait image $\targetimagepred$. At each decoder layer we use a skip connection from the corresponding encoder layer and concatenate the network activations accordingly.

All conv layers are followed by a group normalization~\cite{Wu_2018_ECCV} and an activation function. A sigmoid activation is used in the layer before the target portrait image prediction (as images are assumed to be in $[0, 1]$), a softplus activation~\cite{Dugas2000} is used immediately before the weighted average that predicts the lighting of the source portrait (as confidence must be non-negative), and a PReLU activation~\cite{he2015delving} is used after all other layers.

\subsection{Loss Function and Training}
\label{sec:model-train}

Our model is trained through the minimization of a weighted combination of three loss functions. The first loss minimizes errors between the true target image $\targetimage$ in our dataset and the  target relit image $\targetimagepred$ predicted by our network according to the ``target`` illumination $\targetlight$. The second loss minimizes errors between the predicted source illumination $\sourcelightpred$ and the true source illumination $\sourcelight$.
To evaluate our third ``self-supervision'' loss, 
we modify our network architecture slightly and replace the true target illumination $\targetlight$ that was previously used as input to the decoder stream of the network with the predicted source illumination $\sourcelightpred$ that is produced by the model, and minimize errors between the resulting reconstruction of the source image $\sourceimagepred$ and the input source image $\sourceimage$. The goal of the first loss is straightforward: we would like our model to produce accurate relightings. The next two losses have a more nuanced effect: by ensuring that the model can accurately reproduce its own input image and illumination, we enable the use-case in which our model is used to alter the existing illumination in the scene, rather than replace it entirely with some new illumination. Additionally, we will show that imposing these two secondary loss functions actually improves performance on our primary goal of relighting.

The loss we will use when comparing relit or reconstructed images is the per-pixel $L_1$ distance between our predicted image $\imagepred$ and the true image $\image$, after the background has been masked out:
\begin{equation}\label{equ:imageloss}
    \imageloss\left(\imagepred, \image\right) = \normone{\mask \odot \left( \imagepred - \image\right) },
\end{equation}
where $\mask$ is a per-pixel portrait mask, and $\odot$ is element-wise multiplication. For the illumination prediction, we use the weighted log-$L_2$ distance of \cite{weber2018learning}:
\begin{equation}\label{equ:lightloss}
    \lightloss\left(\lightpred, \light\right) = \normtwo{\solidangle \odot \left( \log(1+\lightpred) - \log(1+\light)\right) }^2,
\end{equation}
where $\solidangle$ is the solid angle of each ``pixel'' in our environment illumination in their latitude-longitude representation.
With these defined, we can construct our complete loss function as:
\begin{align} 
\loss = \imageloss\left(\targetimagepred, \targetimage\right) +  \lambda_{\mathit{light}}\cdot \lightloss\left(\sourcelightpred, \sourcelight\right) +
\lambda_{\mathit{self}} \cdot \imageloss\left(\sourceimagepred, \sourceimage\right),  \label{equ:loss}
\quad\quad
\end{align}
where we set the hyperparameters $\lightweight = 0.8$ and $\selfweight=1$.

As is often the case with autoencoders, our self-supervision loss works best if the predicted source illumination that is fed back into the decoder of the network is ``jittered'' slightly. Otherwise the network may begin to generalize poorly to illuminations that are similar to training data instances but are not themselves present in the data, which results in flickering when synthetically rotating the illumination during relighting and in color shifts when relighting in general. To ameliorate this, during training we apply a random rotation of $\theta \in [0^\circ, 360^\circ]$ to each environment along its longitude, and correspondingly re-render the ``source'' image with that rotated environment.

Our model is implemented using TensorFlow~\cite{Tensorflow}. We train our model on 4 NVIDIA Tesla V100s with batch size 2 on each device. We use the Adam optimizer of \cite{KingmaB14} with a learning rate of $10^{-5}$ to train the network, and our model converges after 20 epochs (which takes $\sim\!26$ hours).

\section{Evaluation}\label{sec:eval}

Here we evaluate our proposed model against prior works and against ablations of our own model. We present results on two tasks: in Section~\ref{sec:eval-image} we measure the quality of the relit images produced by each model, and in Section~\ref{sec:eval-light} we measure the quality of the environmental illuminations produced by each model.

\newcommand{\scalemse}{\operatorname{RMSE-s}}

\definecolor{Yellow}{rgb}{1,1, 0.6}
\definecolor{Orange}{rgb}{1,0.8, 0.6}
\definecolor{Red}{rgb}{1, 0.6, 0.6}

\begin{table*}[]
\caption{
Here we benchmark our model on the validation set of our dataset against prior work on single-image relighting and against ablations of our model. The ``Target'' metrics measure each algorithm's accuracy in predicting each desired relit image, and the ``Source'' metrics measure each algorithm's accuracy in predicting the input image itself according to the model's predicted illumination.
The reported metrics and runtimes are the arithmetic mean over all images in the validation set.
The top three techniques for each metric are highlighted in red, orange, and yellow respectively.
\vspace*{-.1in}
\label{table:compare}}
\begin{center}
\begin{tabular}{ l || c c c || c c c || c }
& \multicolumn{3}{c||}{Target} & \multicolumn{3}{c||}{Source} & \\
Algorithm & RMSE & $\scalemse$ & DSSIM & RMSE & $\scalemse$ & DSSIM & Time (sec.) \\
\hline
SIRFS~\cite{BarronTPAMI2015} & 0.1715 & 0.0935 & 0.0827 & \cellcolor{Red} 0 & \cellcolor{Red} 0 & \cellcolor{Red} 0 & 788 \\
SfSNet~\cite{sfsnetSengupta18}& \changed{0.2397} & \changed{0.0997} & \changed{0.1297 }& 0.0698 & 0.0673 & 0.0899 & 0.7442 \\
\hline
Our model & \cellcolor{Red} 0.0435 & \cellcolor{Red} 0.0351 & \cellcolor{Red} 0.0251 & \cellcolor{Orange} 0.0170 & \cellcolor{Orange} 0.0156 & \cellcolor{Orange} 0.0060 & 0.1616 \\
Ours w/o light prediction  & \cellcolor{Yellow} 0.0460 & \cellcolor{Yellow} 0.0395 & \cellcolor{Red} 0.0251 & - & - & 
- & - \\
Ours w/o self-supervision & \cellcolor{Orange} 0.0448 & \cellcolor{Orange} 0.0370 & \cellcolor{Yellow} 0.0258 & \cellcolor{Yellow} 0.0313 & \cellcolor{Yellow} 
0.0282 & \cellcolor{Yellow} 0.0118 & - \\
\end{tabular}
\end{center}
\vspace*{-.1in}
\end{table*}

\subsection{Relit Images}\label{sec:eval-image}
{\bf Metrics: } We measure the relighting performance using three error metrics across our validation set: RMSE, scale-invariant RMSE ($\scalemse$), and DSSIM.
Our scale-invariant RMSE solves for the single scale-factor that can be applied to the predicted image to best minimize RMSE.
\begin{equation}
\scalemse( \targetimagepred, \targetimage) = \min_\alpha \normtwo{ \alpha\targetimagepred - \targetimage }
\end{equation}
We use this metric because a relit image that is correct up to a scale factor may still be useful in many photographic contexts. Note that this metric solves for a single global scaling rather than a per-channel scaling, meaning that it still sensitive to erroneous tints in the output image.
Our DSSIM implementation uses a $11 \times 11$ Gaussian filter with $\sigma=1.5, k1 = 0.01, k2 = 0.03$, as is recommended by \cite{SSIM}.
DSSIM is computed on each RGB channel individually, which (because DSSIM is invariant to local scaling) means that our DSSIM metric is invariant to both global and local scaling and color shifts.
Our metrics thereby span a continuum of invariance, from RMSE (which is invariant to nothing) to DSSIM (which is invariant to local and global scaling and tinting).
This continuum is worth considering when designing error metrics for this task, as a common mistake for relighting or intrinsic image decomposition algorithms is that of erroneously producing a scaled up illumination and a correspondingly scaled down albedo.

\newcommand{\resultswidth}{0.164\textwidth} 
\setlength{\tabcolsep}{0pt} 
\begin{figure*}[!ht]
  \centering
\begin{tabular}{cc|cccc}
& & \multicolumn{4}{c}{Target image prediction} \\
(a) Source image & (b) Target image & (c) Our model &
(d) \footnotesize{\cite{BarronTPAMI2015}} &
(e) \small{\cite{sfsnetSengupta18}} &
(f) \small{\cite{li2018learning}}\\
\includegraphics[width=\resultswidth]{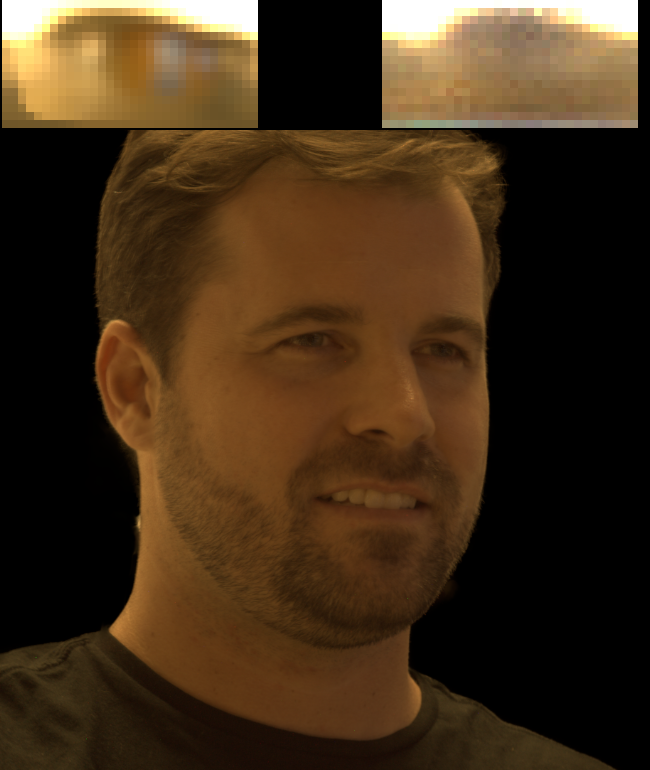}&
\includegraphics[width=\resultswidth]{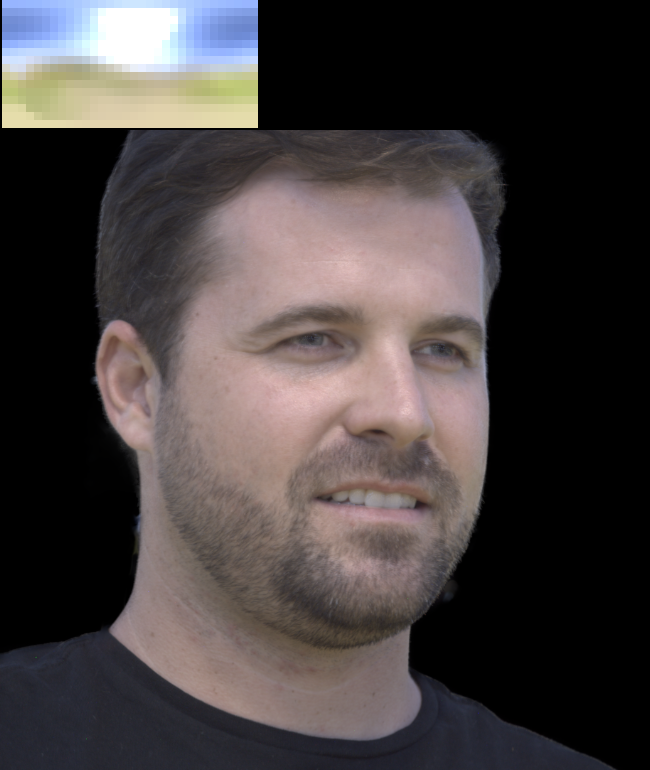}&
\includegraphics[width=\resultswidth]{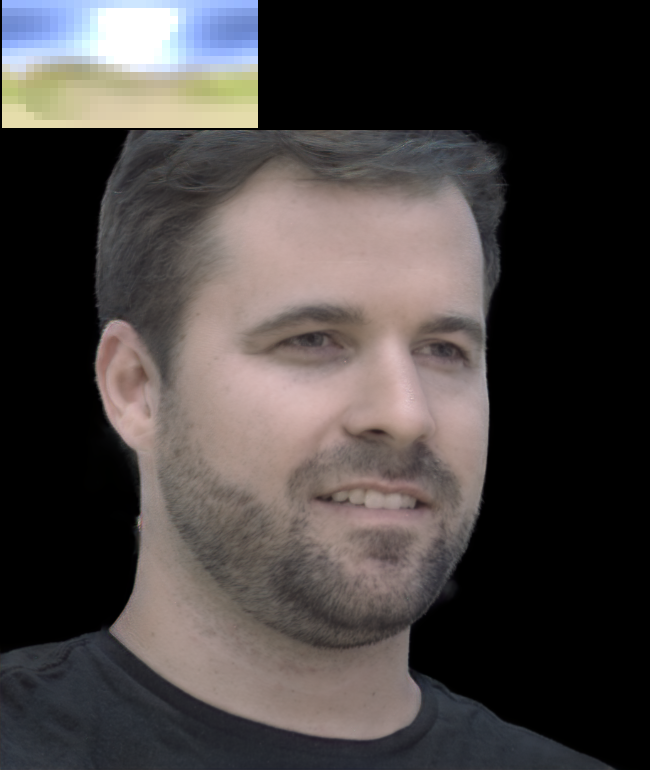}&
\includegraphics[width=\resultswidth]{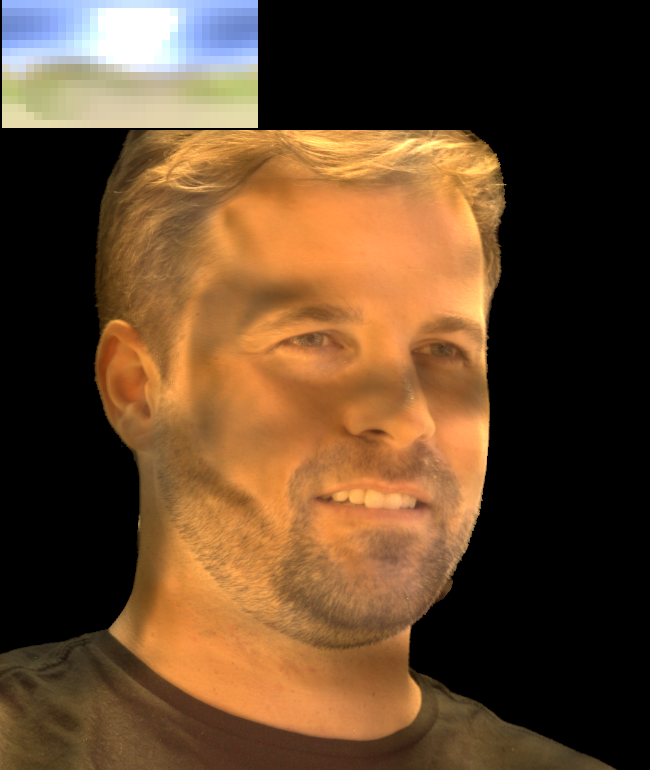}&
\includegraphics[width=\resultswidth]{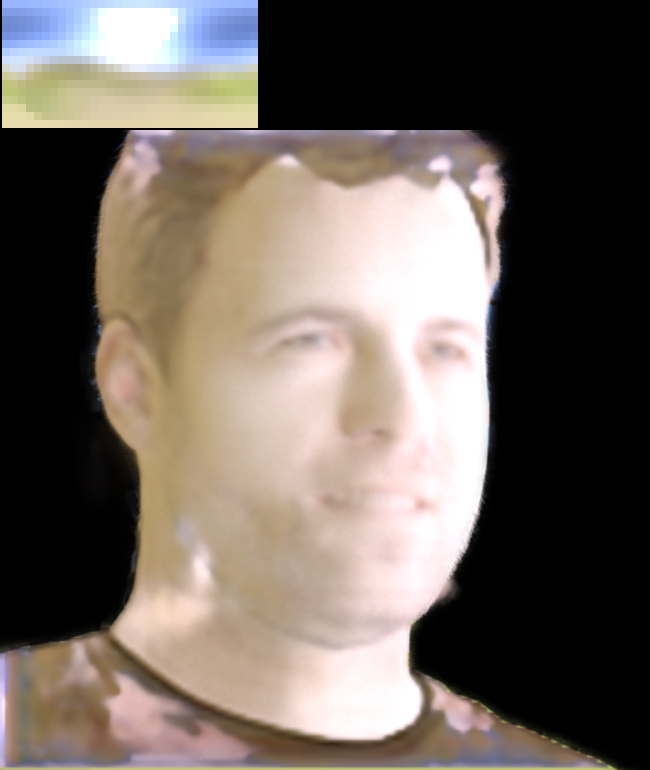}&
\includegraphics[width=\resultswidth]{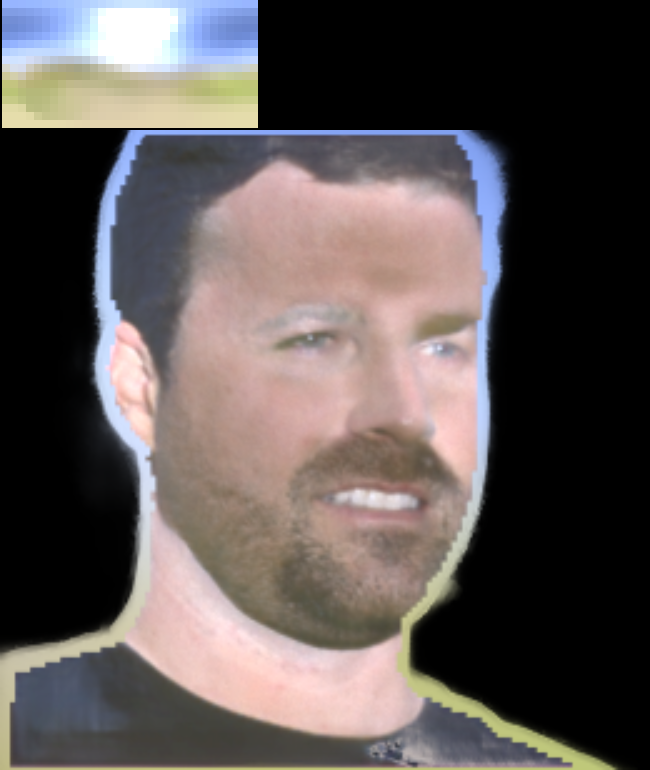}\\
\includegraphics[width=\resultswidth]{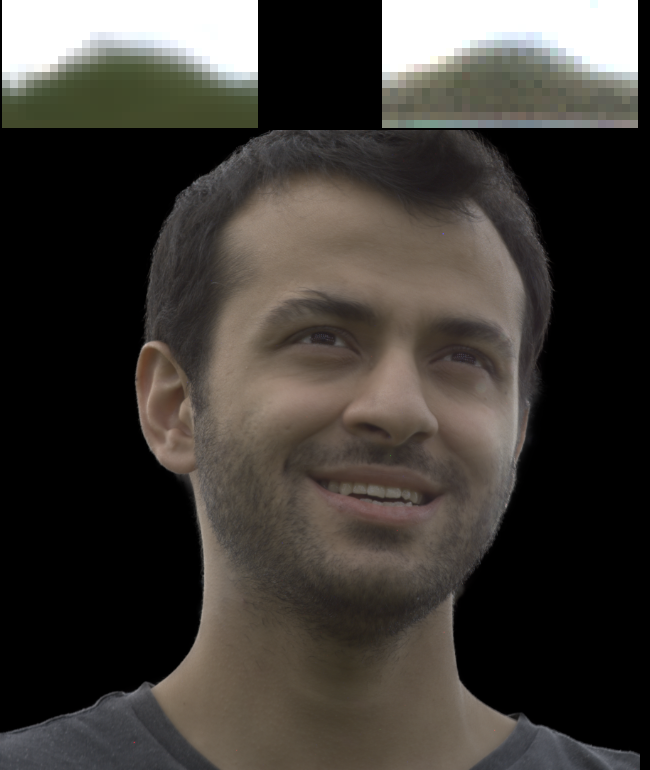}&
\includegraphics[width=\resultswidth]{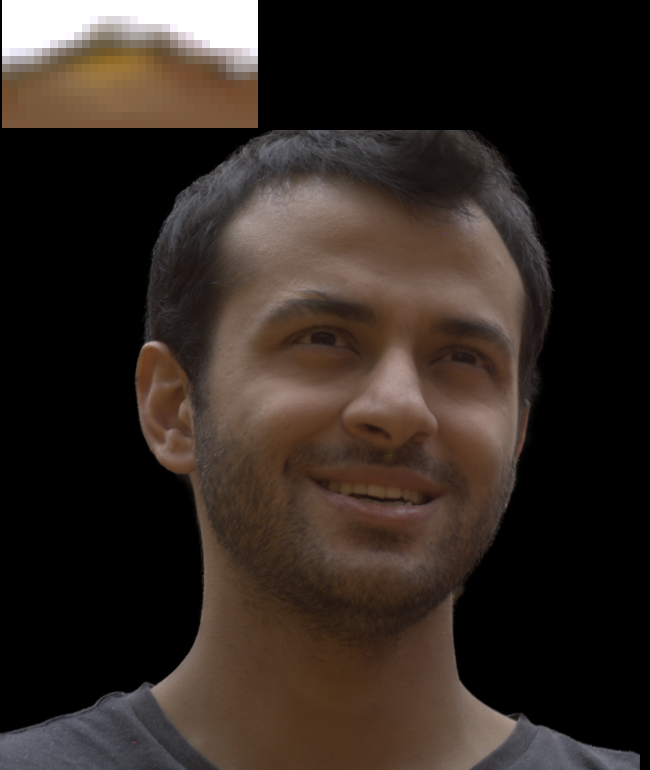}&
\includegraphics[width=\resultswidth]{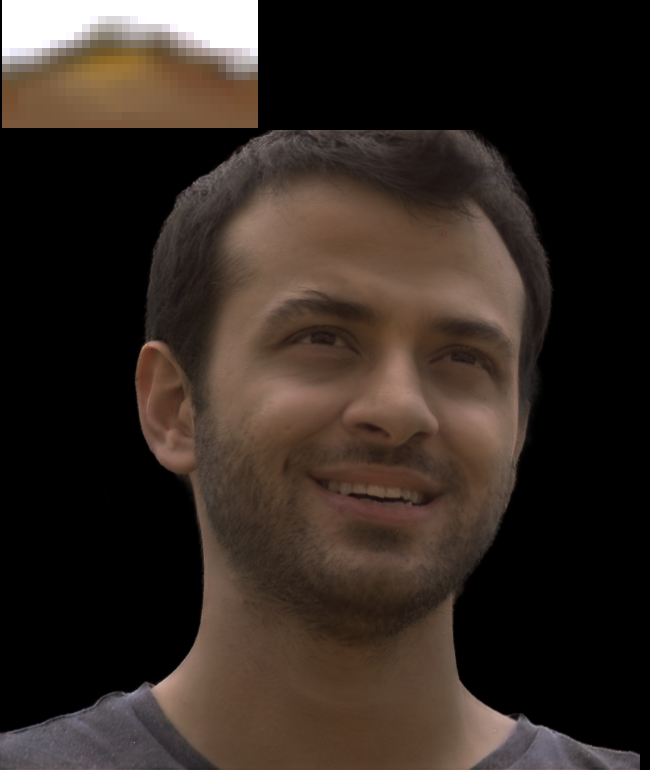}&
\includegraphics[width=\resultswidth]{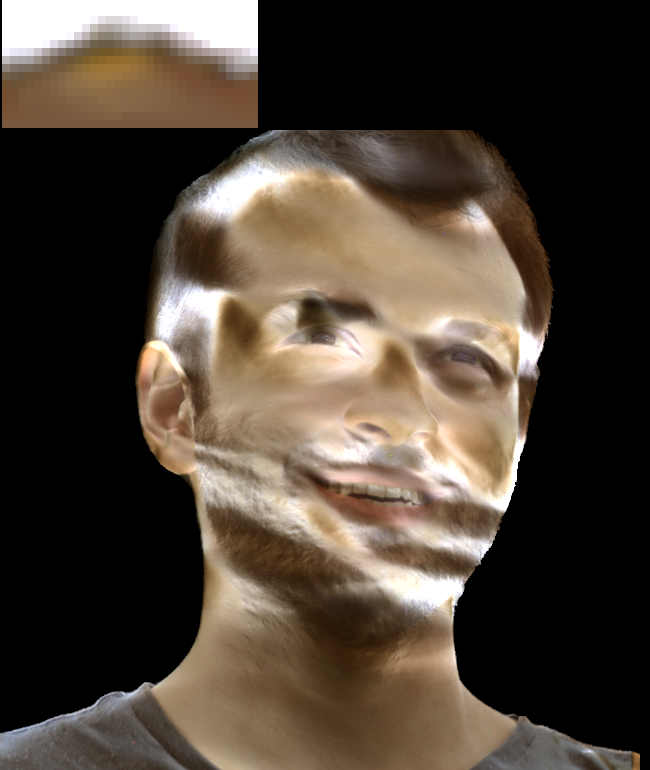}&
\includegraphics[width=\resultswidth]{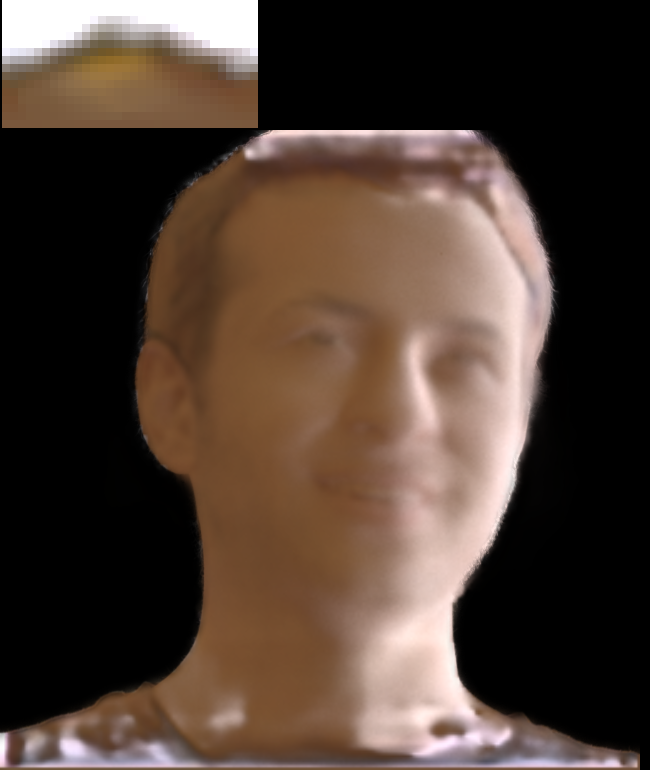}&
\includegraphics[width=\resultswidth]{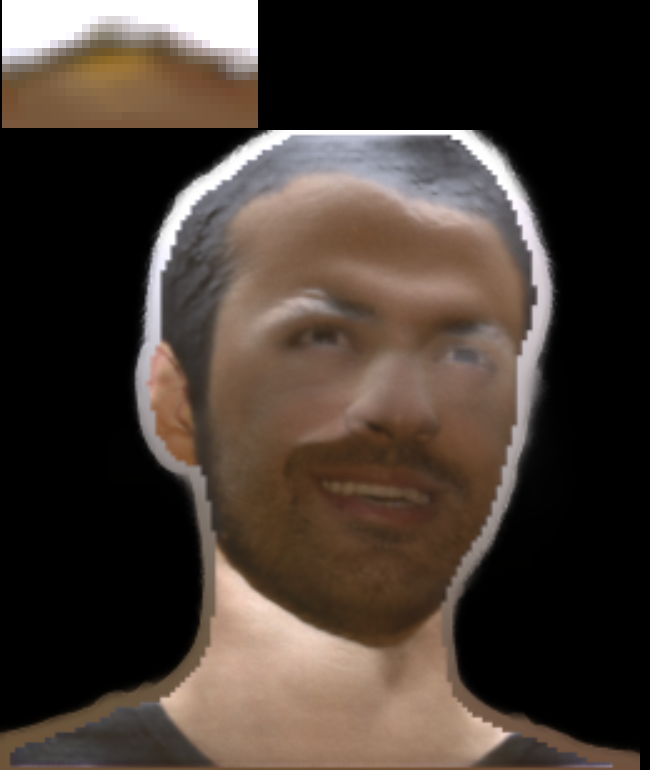}\\
\includegraphics[width=\resultswidth]{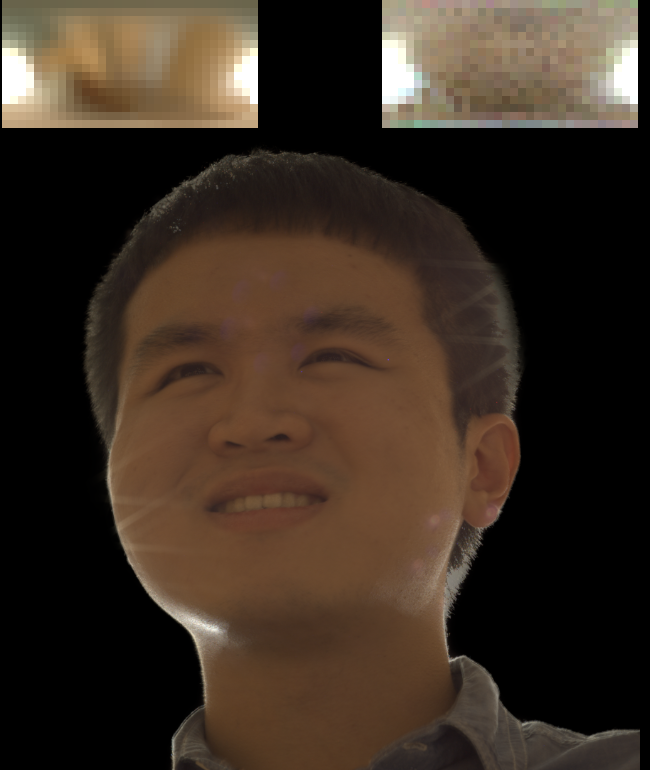}&
\includegraphics[width=\resultswidth]{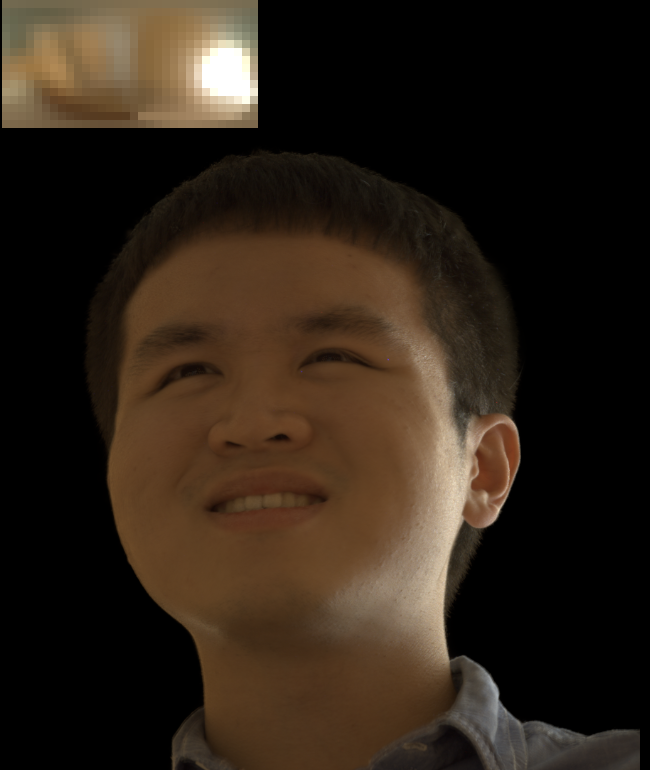}&
\includegraphics[width=\resultswidth]{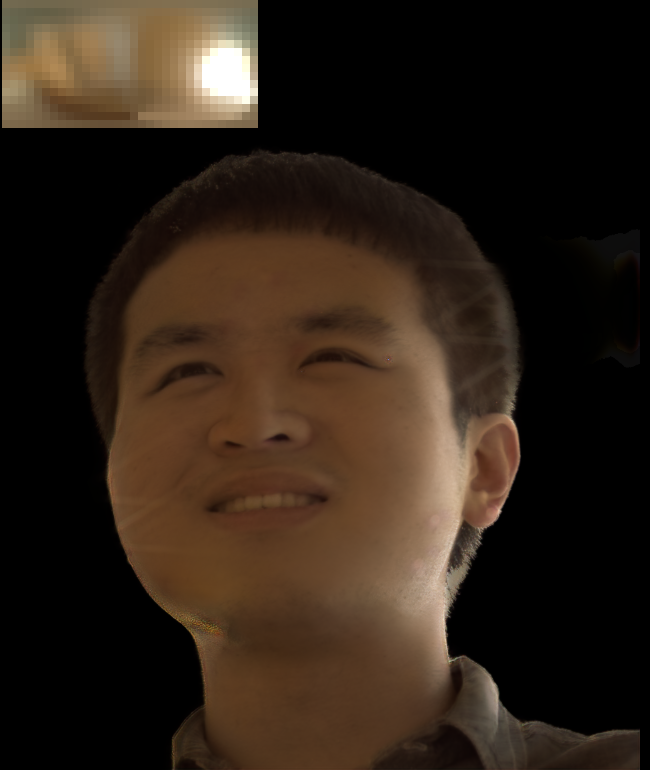}&
\includegraphics[width=\resultswidth]{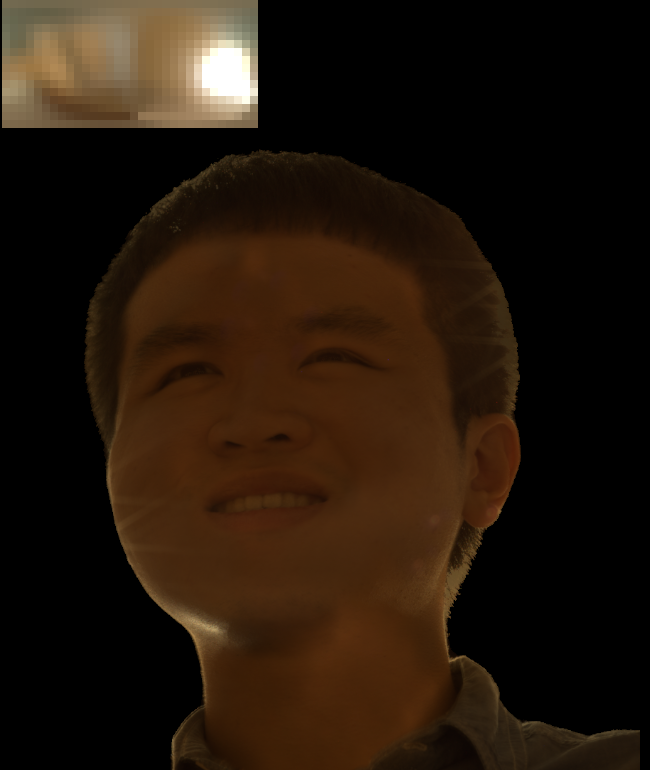}&
\includegraphics[width=\resultswidth]{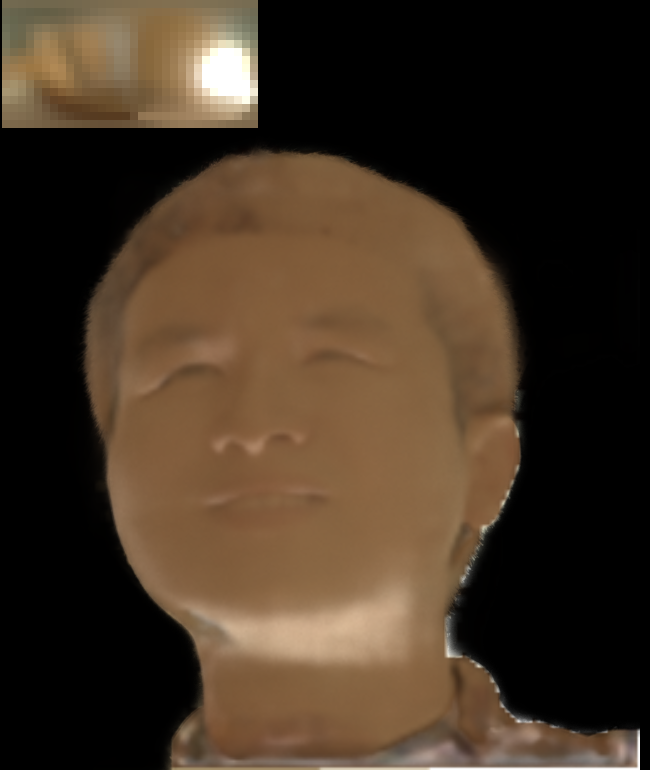}&
\includegraphics[width=\resultswidth]{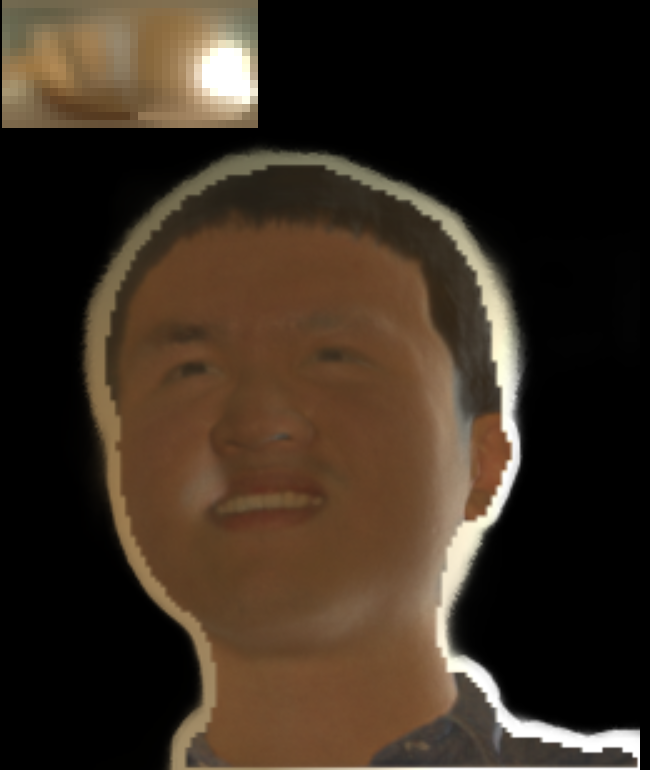}\\
\end{tabular}
\vspace*{-.1in}
\caption{
Here we present results using the validation set of our dataset. The ``Source image'' (a) shows the input image to our system and the true illumination (top left) that lit this scene, as well as the illumination predicted by our model (top right). The ``Target image'' (b) shows the true illumination (top left) that we wish to relight the ``source'' image with, as well as the true appearance of the subject under that illumination. The remaining columns compare the output of our model (c) with three state-of-the-art baseline models (d, e, f).
Our model's output is significantly more realistic and compelling than that of any baseline, even when presented with challenging scenarios such as relighting from a back-lit scene to a front-lit scene (row 1), cast shadows (row 2), and subsurface-scattering and translucency due to backlighting (row 3).}
\label{fig:compare}
\vspace*{-.1in}
\end{figure*}
\setlength{\tabcolsep}{6pt} 

\renewcommand{\resultswidth}{0.157\textwidth}
\setlength{\tabcolsep}{0pt} 
\begin{figure*}[!ht]
  \centering
\begin{tabular}{@{}c@{}c@{\,\,}c@{\,\,}|@{\,\,}c@{}c@{}c}
& & & \multicolumn{3}{c}{Output target image prediction} \\
(a) Source image & (b) Target image & (c) Reference image &
(d) Our model &
(e) {\cite{Shih:2014:STH}} &
(f) {\cite{shu2018portrait}}\\
\includegraphics[width=\resultswidth]{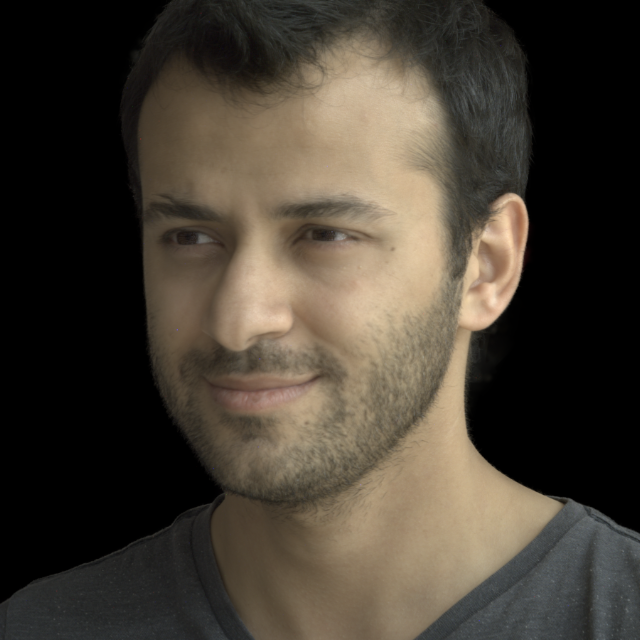}&
\includegraphics[width=\resultswidth]{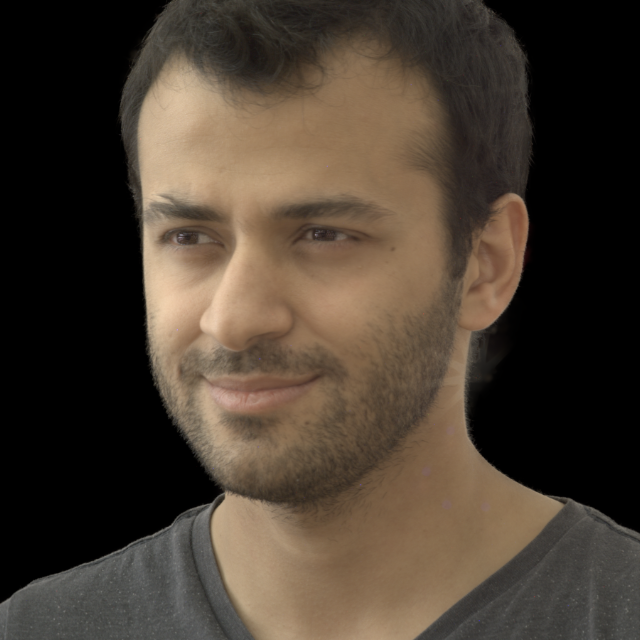}&
\includegraphics[width=\resultswidth]{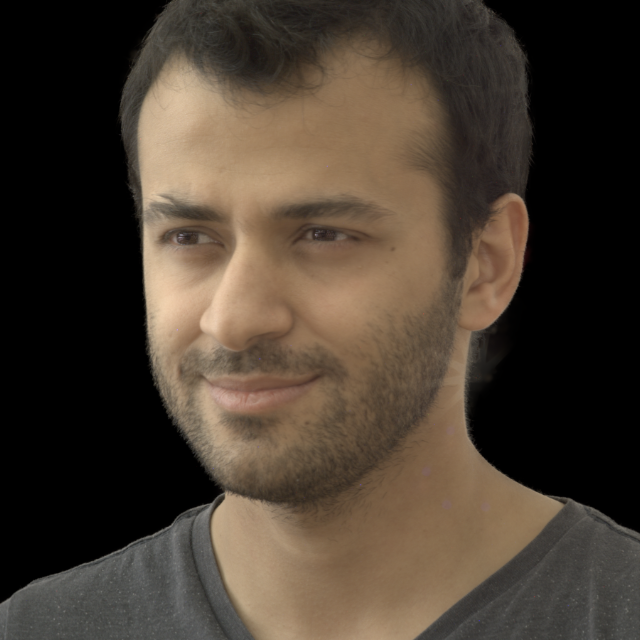}&
\includegraphics[width=\resultswidth]{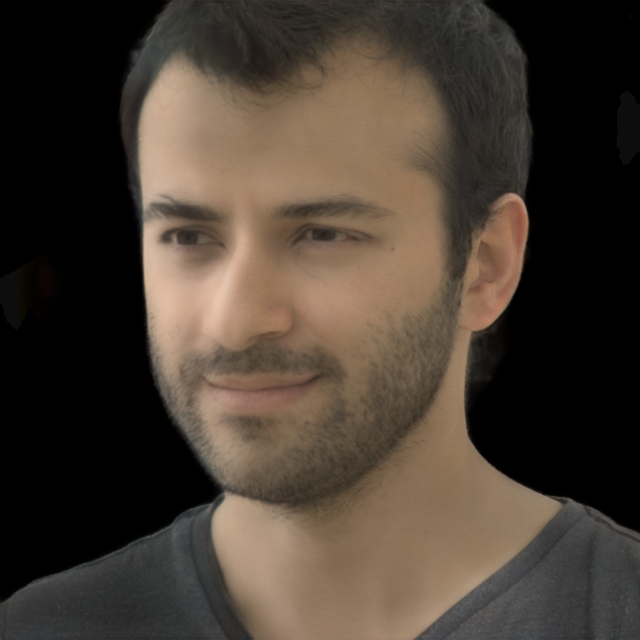}&
\includegraphics[width=\resultswidth]{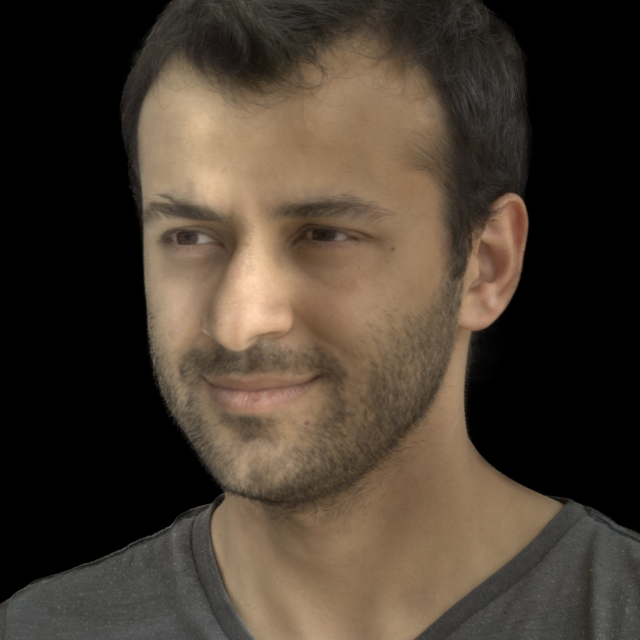}&
\includegraphics[width=\resultswidth]{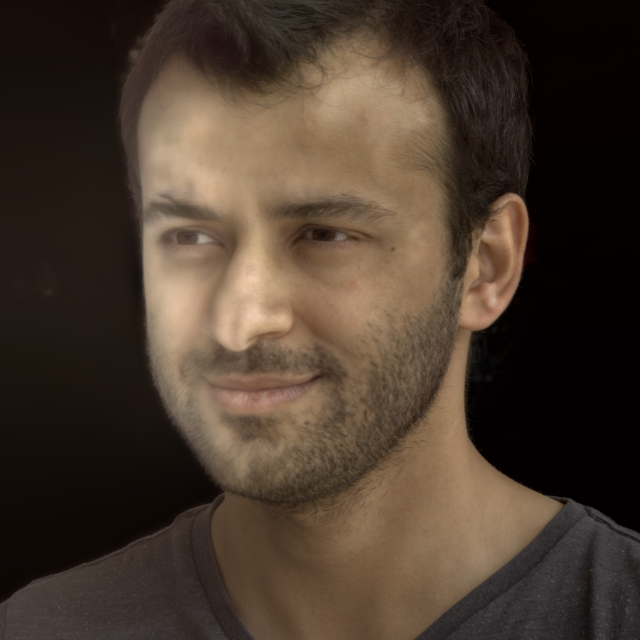}\\
\includegraphics[width=\resultswidth]{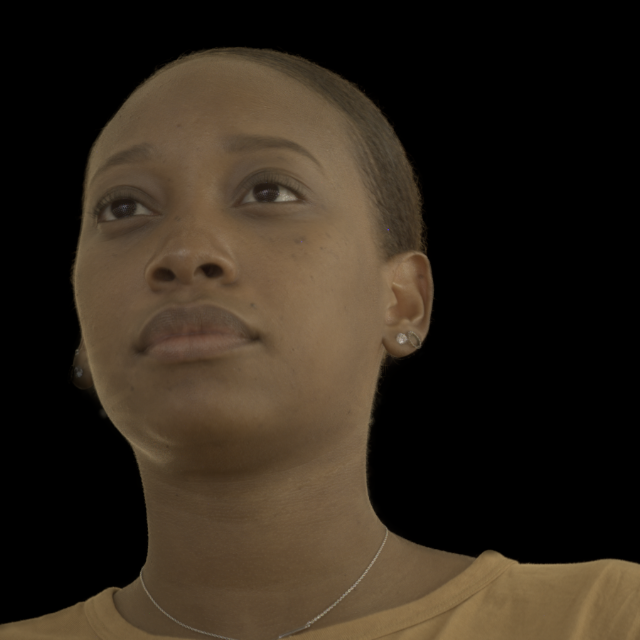}&
\includegraphics[width=\resultswidth]{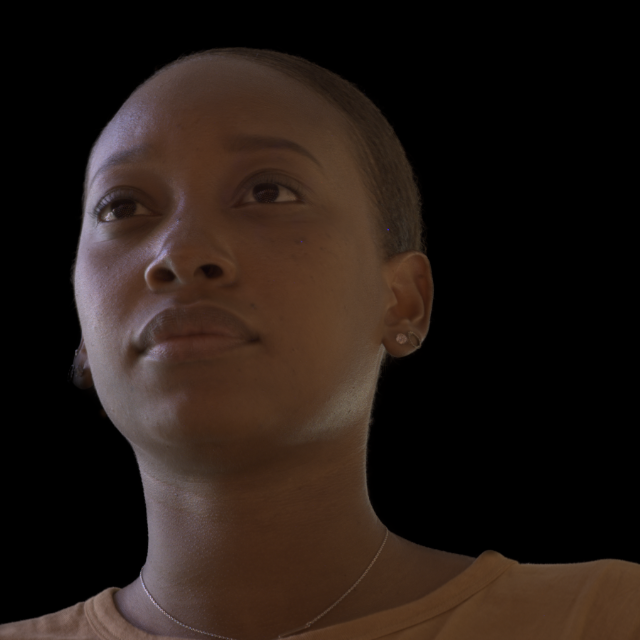}&
\includegraphics[width=\resultswidth]{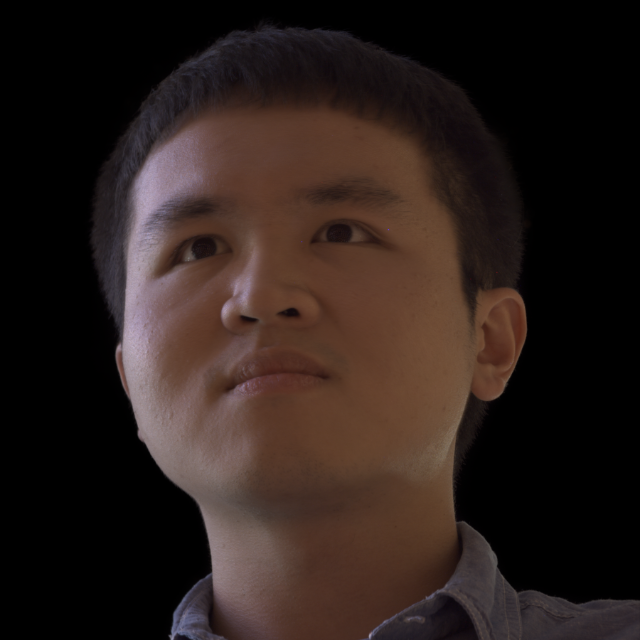}&
\includegraphics[width=\resultswidth]{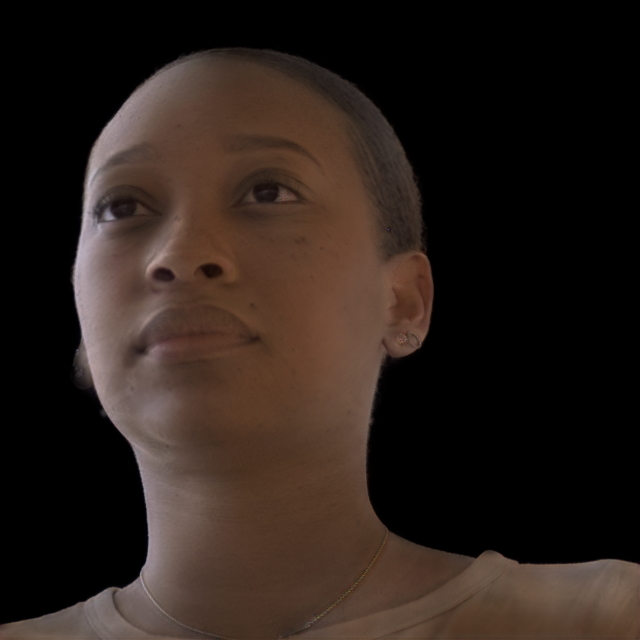}&
\includegraphics[width=\resultswidth]{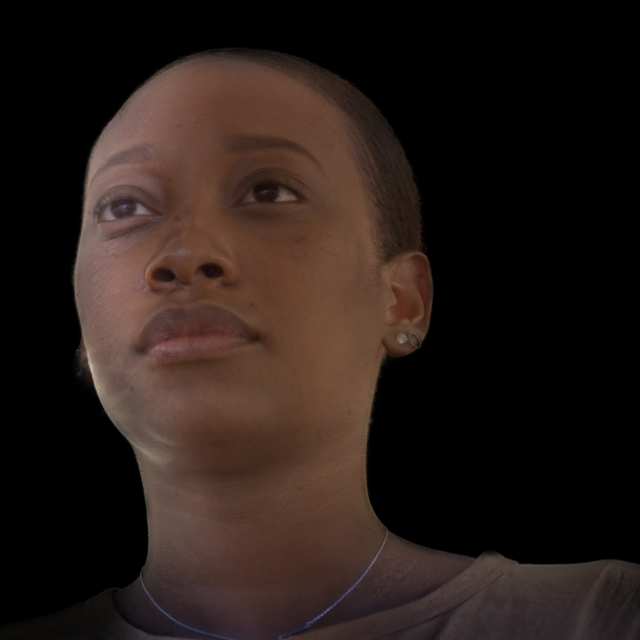}&
\includegraphics[width=\resultswidth]{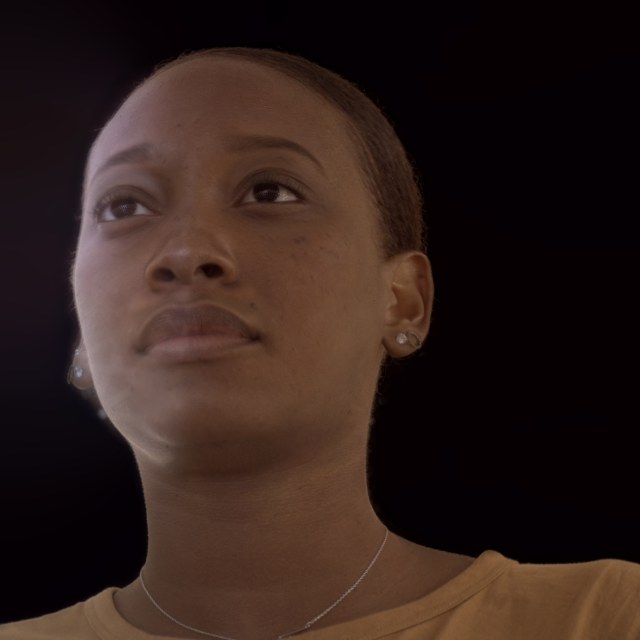}
\end{tabular}
\vspace*{-.1in}
\caption{\changed{Here we show a qualitative comparison between lighting transfer algorithms our method. In the top row we provide the target image as a reference to each lighting transfer algorithm, and in the bottom row we provide a random subject as a reference (these reference images are not used by our model).
Under both setups, our model produces a more natural lighting on the portrait, while both baseline lighting transfer methods introduce artifacts on human faces.}}
\label{fig:light_transfer}
\vspace*{-.1in}
\end{figure*}
\setlength{\tabcolsep}{6pt} 
\renewcommand{\arraystretch}{1} 

{\bf Intrinsic decomposition methods: } In Table~\ref{table:compare} we present the performance of our models compared to several state-of-the-art methods for the single-image relighting task on the validation set of our dataset. This table primarily serves to report the performance of each algorithm in terms of prediction of the ``target'' portrait image for each validation set scene, though we additionally report the accuracy of each model in reconstructing the source image --- the input image itself --- according to the predicted illumination from the model. These ``source'' error metrics are of interest for the use-case in which a user of our system wishes to modify the illumination of the input image, rather than completely replace it.
Note that it is possible for a model to trivially minimize the ``source'' error metrics by simply returning the input image, but it is not possible to trivially minimize the ``target'' error metrics.
In addition to these error metrics, we additionally report the average runtimes for each technique across the validation set, all of which were benchmarked using the same computer that was used during training.

Our comparison against SIRFS~\cite{BarronTPAMI2015} was done using the code provided by the authors, with no modification. Relighting is performed by taking the surface normals and reflectance recovered by the model along with the desired environment illumination, then solving a linear system to recover the spherical harmonic illumination that best reproduces the log of the environment illumination (this model is designed to work with ``log-shading''), and then rendering the surface normals and reflectance according to that spherical harmonic illumination.
Note that the source error metrics are trivially minimized by this baseline, because the intrinsic decomposition produced by this model is constrained to exactly reproduce the input image by construction. Because the SIRFS supports arbitrary-resolution inputs, images are processed at their native resolution, and the output relit images are downsampled to the $640 \times 640$ resolution used by our model. 
Our comparison against the SfSNet~\shortcite{sfsnetSengupta18} was also done using the code provided by the authors with no modification, and relighting was performed by rendering the output normals and reflectance produced by the model with \changed{its predicted environment illumination}. Because this model only supports $128 \times 128$ resolution inputs, we downsample the input image before relighting and then upsample the output relit image to $640 \times 640$ before computing our error measures.

From Table~\ref{table:compare} we see that our model outperforms prior work in terms of all three of our ``Target'' error metrics by a significant margin, with reduced error rates compared to the next best-performing technique by $57\%$ - $75\%$. Our performance on the ``Source'' error metrics is the second lowest after the SIRFS~\cite{BarronTPAMI2015}, which by construction perfectly reproduces its input image.
Additionally, our technique is $\sim\!4.6\times$ faster than the SfSNet~\cite{sfsnetSengupta18} and $\sim\!4800 \times$ faster than the SIRFS~\cite{BarronTPAMI2015}.
To further demonstrate this improved performance, in~\fig{fig:compare} we visualize our model's performance compared to our baselines on three of our validation set scenes, and see that our model consistently produces realistic relighting results while the baseline techniques do not.
\fig{fig:compare} also includes a comparison against the technique of Li \etc~\shortcite{li2018learning}, which (unlike our technique and our primary baselines) requires an input image in which the illumination is fronto-parallel. To produce these baseline results we use the fronto-parallel illuminated image from each OLAT imageset as input to this model, thereby satisfying its constraints. This is a generous comparison, as a known and controlled illumination significantly reduces the difficulty of single-image relighting. Despite this advantage, as shown in \fig{fig:compare}, this approach does not produce satisfactory results on this task.

\begin{table}[]
\caption{\changed{
Here we compare our model against two lighting transfer algorithms. 
Each transfer algorithm is evaluated in two settings: a generous setting one in which the ground-truth output image is presented as input as a reference, and a more-fair setting in which the reference image is a portrait of a different subject illuminated by the desired lighting.
Our model significantly outperforms the two baselines in the randomized setting, and even outperforms the two baselines on two of our three metrics in the ``oracle'' setting where the baselines have access to the ground-truth}
\vspace*{-.15in}
\label{table:light_transfer}}
\begin{center}
\resizebox{\linewidth}{!}{%
\begin{tabular}{ l || c c c }
& \multicolumn{3}{c}{Target} \\
Algorithm & RMSE & $\scalemse$ & DSSIM \\
\hline
\cite{Shih:2014:STH} + ground-truth & \cellcolor{Red} 0.0406 & \cellcolor{Orange} 0.0401 & 0.0373  \\
\cite{shu2018portrait} + ground-truth  & 0.0563 & 0.0528 & 0.0408\\
\hline
\cite{Shih:2014:STH} & 0.1012 & 0.0874 & 0.0653  \\
\cite{shu2018portrait} & 0.0697 & 0.0456 & \cellcolor{Orange} 0.0300  \\
\hline
Our model & \cellcolor{Orange}    0.0435 & \cellcolor{Red}    0.0351 & \cellcolor{Red}    0.0251  \\
\end{tabular}
}
\end{center}
\vspace*{-.1in}
\end{table}

\newcommand{\reconwidth}{0.142\textwidth}
\setlength{\tabcolsep}{0pt} 
\begin{figure}[!h]
  \centering
\begin{tabular}{ccc}
{\small Source image}&
{\small Our model}&
{\small \shortstack{Ours w/o \\ self-supervision}}\\
\includegraphics[width=\reconwidth]{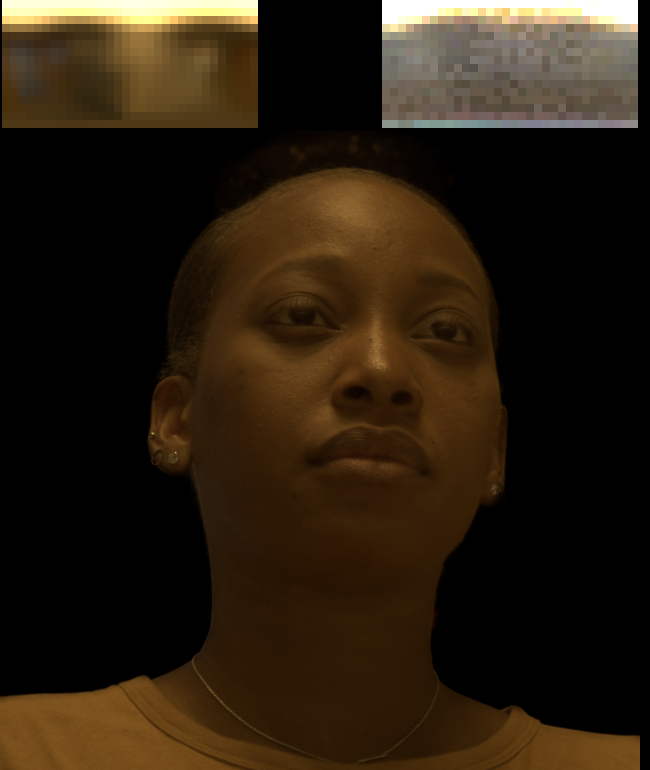}&
\includegraphics[width=\reconwidth]{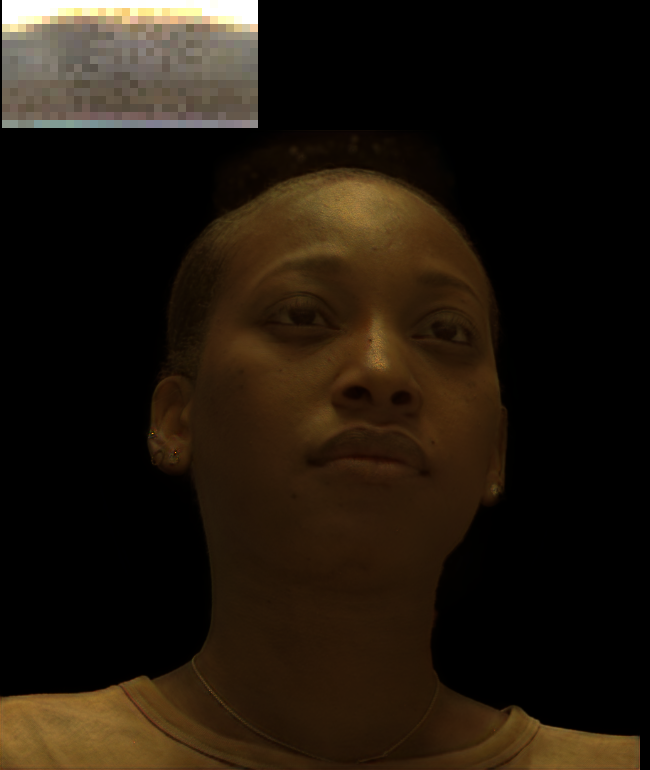}&
\includegraphics[width=\reconwidth]{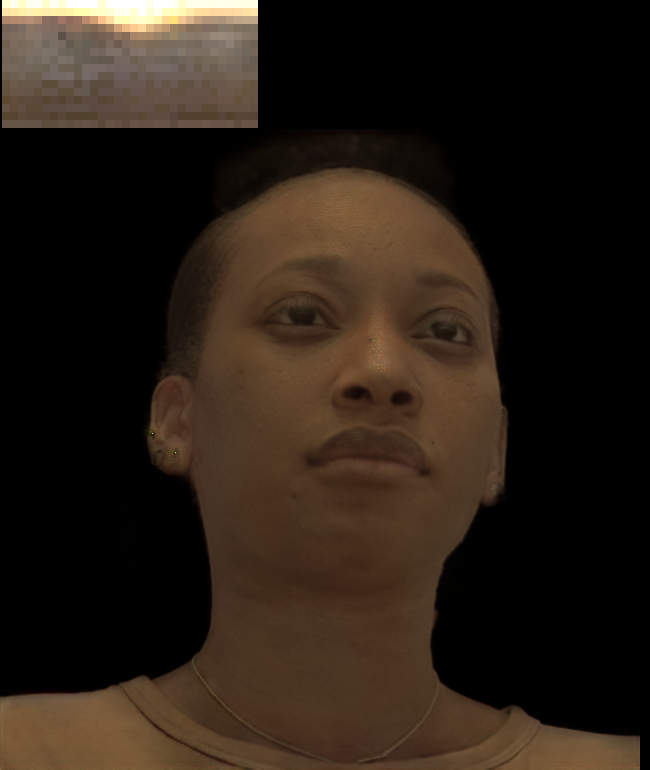}\\
\includegraphics[width=\reconwidth]{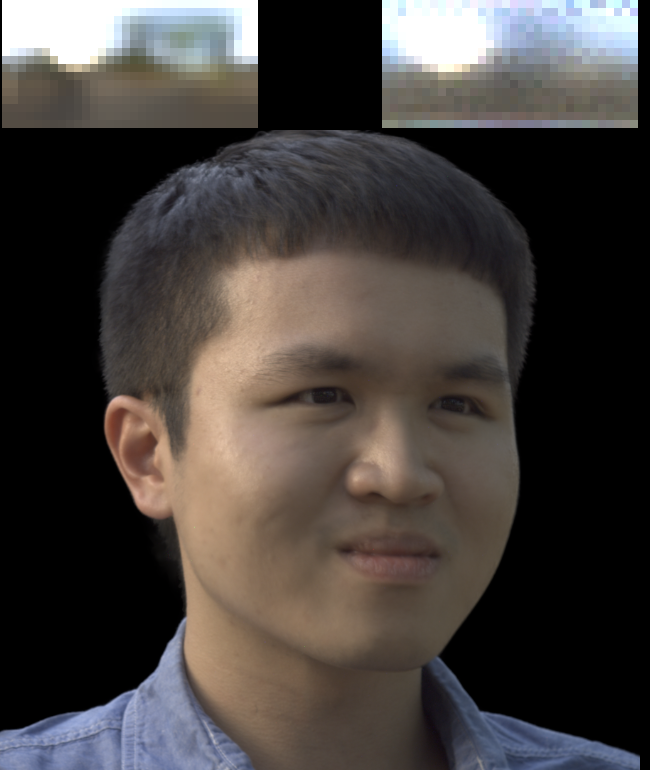}&
\includegraphics[width=\reconwidth]{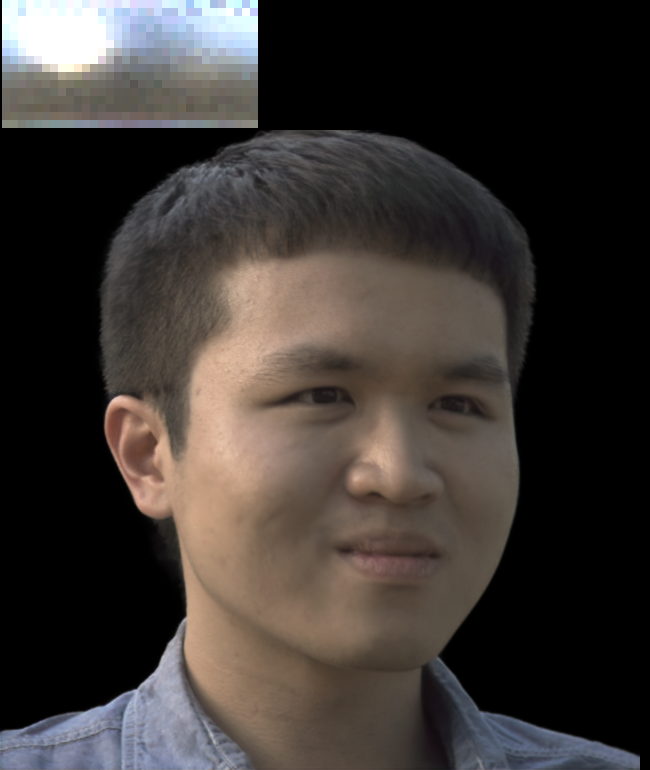}&
\includegraphics[width=\reconwidth]{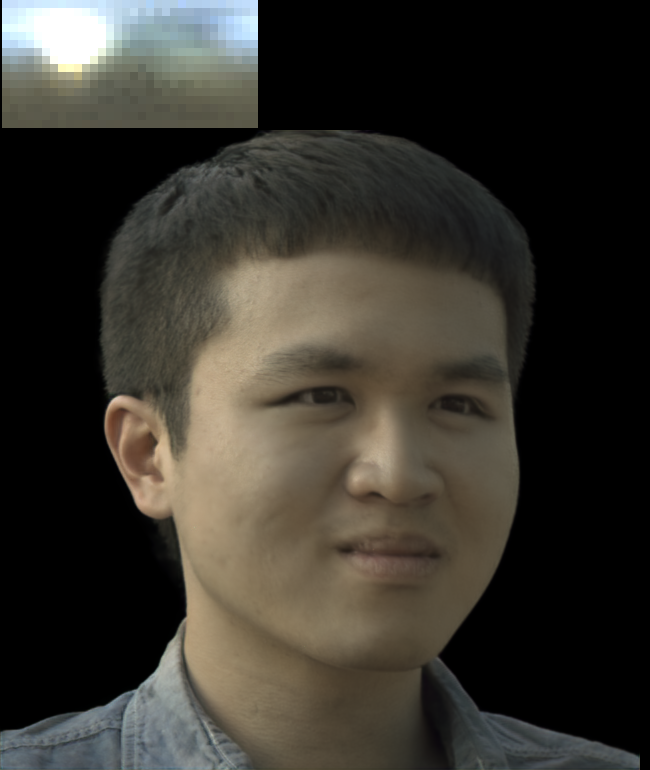}\\
\includegraphics[width=\reconwidth]{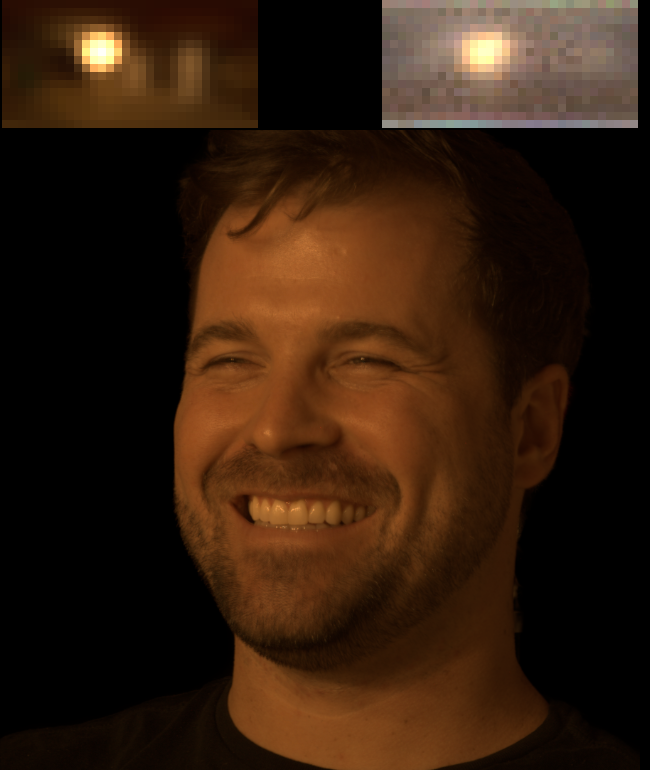}&
\includegraphics[width=\reconwidth]{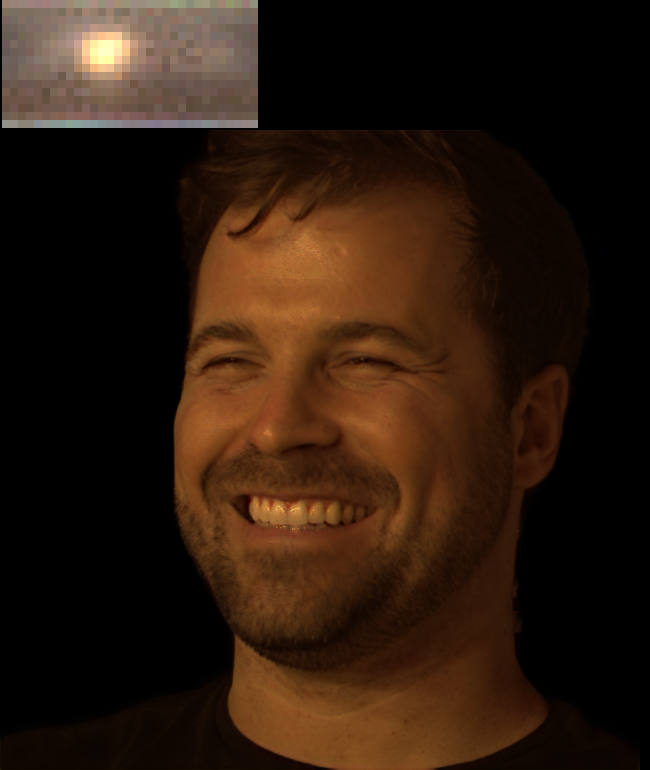}&
\includegraphics[width=\reconwidth]{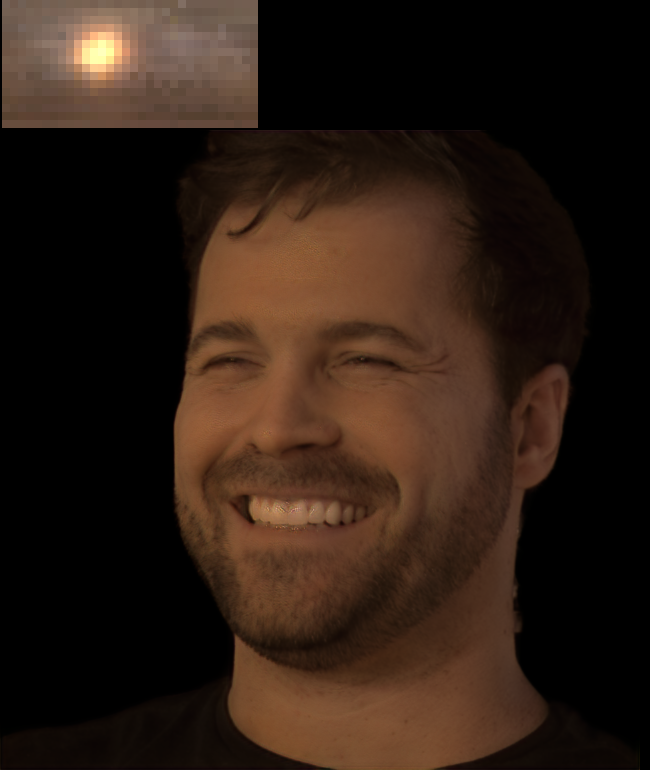}
\end{tabular}
\vspace*{-.05in}
\caption{
Instead of completely replacing the true illumination of the scene (left top inset), one may want to instead recover the existing illumination of the scene (right top inset) and then modify it.
To this end, we impose a self-supervision loss during training, which causes our model (middle column) to more accurately reproduce the input (left column) than a model trained without self-supervision (right column). Self-supervision helps our model correctly disentangle skin color from illumination color, as evidenced by the unnatural skin colors in the model without self-supervision.
}
\label{fig:ablation}
\vspace*{-.1in}
\end{figure}
\setlength{\tabcolsep}{6pt} 
\renewcommand{\arraystretch}{1} 

{\bf Lighting transfer methods: } 
\changed{
In Table~\ref{table:light_transfer} we compare our model against two illumination \emph{transfer} algorithms~\cite{Shih:2014:STH,shu2018portrait}, which do not perform relighting in the same manner as our model, but instead work by transferring the lighting conditions from one portrait to another. These transfer algorithms take as input two images: the ``source'' image that is to be modified, and a ``reference'' image that contains the illumination that we would like to transfer. The output of these models should be the subject of the source image lit by the illumination in the reference image (the ``target'' image, using the nomenclature of this work).}

\changed{
Comparing our model against these transfer techniques is challenging, as they do not learn from training data, and instead require a reference image as input. Therefore, when evaluating against these models on our validation set, we must decide which image from our training data to present to each model as its ``reference''.
In an attempt to produce a fair comparison, we primarily evaluate against each baseline by using a random image of a different person rendered under the ``target'' illumination as the reference image.
In this setting, we outperform both baselines on all error metrics by a large margin.
We also present additional ``oracle'' results in which we use the ground-truth target image directly as the reference image.
This comparison is generous to the transfer techniques, as simply returning the reference image would result in a perfect reconstruction of the target image with zero error. 
Performance in this ``oracle'' setting should be thought of as a bound on performance in more realistic settings.
Despite the advantage of the transfer techniques in this setting, our model still outperforms these two ``oracle'' baseline techniques in two of the three error metrics we evaluate against.
See Figure~\ref{fig:light_transfer} for a qualitative comparison of our model and these lighting transfer algorithms.}

{\bf Ablation study: } We additionally perform an ablation study of our own model to demonstrate the contribution of specific model components.
In the model labeled ``Ours w/o Light Prediction'' in Table~\ref{table:compare} we set $\lightweight = 0$, thereby disabling the loss on the branch of our network architecture that would otherwise predict the illumination of the input image. As such, this model is unable to reconstruct its own input image (hence the missing ``Source'' error metrics for this baseline), but we also see that this model performs worse on the ``Target'' metrics than our complete model, thereby demonstrating this additional illumination supervision improves performance on the relighting task.
In the model labeled ``Ours w/o Self-Supervision'' we set $\selfweight=0$, thereby disabling the self-supervision loss that would otherwise encourage training to accurately reconstruct the source input image.
This decreases the accuracy of the model in terms of the source error metrics, as one would expect, but it also reduces performance in terms of the target error metrics, thereby demonstrating that self-supervision is a useful cue even if accurately reproducing the input image is not a desired goal of the model. See \fig{fig:ablation} for a visualization of the effect of self-supervision.

\newcommand{\lightwidth}{0.125\textwidth}
\begin{figure}[t]
  \centering
\begin{tabular}{c|cc}
Input image & \multicolumn{2}{c}{Estimated lighting} \\
\multirow{4}{*}[-7pt]{
\shortstack{\includegraphics[width=0.15\textwidth]{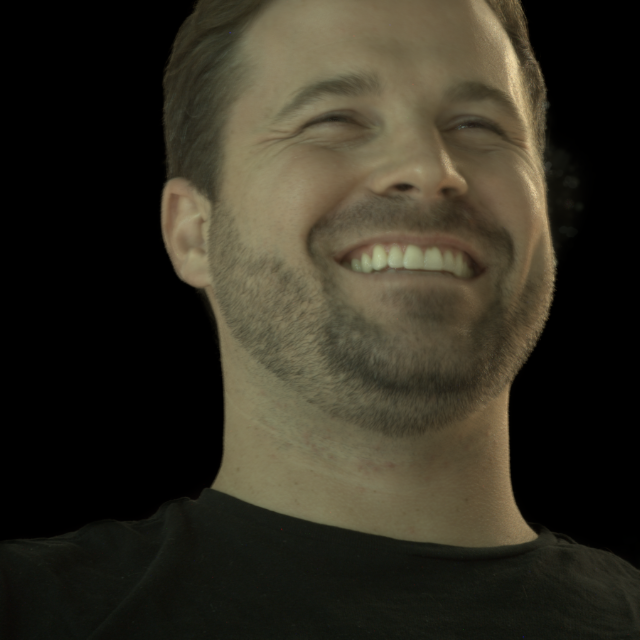}\\
\raisebox{-0.3cm}[0pt][0pt]{\hspace*{-0.9cm}\includegraphics[width=0.1\textwidth]{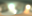}}
}
}
& {\footnotesize Our model} & {\footnotesize \cite{BarronTPAMI2015}}\\
&\includegraphics[width=\lightwidth]{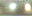}&
\includegraphics[width=\lightwidth]{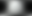}\\
& {\footnotesize \shortstack{Ours w/o\\confidence learning}}& {\footnotesize \cite{sfsnetSengupta18}}\\
&\includegraphics[width=\lightwidth]{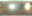}&
\includegraphics[width=\lightwidth]{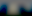}\\
\end{tabular}
\vspace*{-.1in}
\caption{\changed{Here we show lighting estimation results for our model and prior works given a single portrait image. Our model can accurately recover the locations and the colors of multiple light sources, while other methods struggle with high-frequency illumination effects or the non-Lambertian properties of human skin.}
\vspace*{-.05in}
\label{fig:light_estimate}}
\end{figure}
\setlength{\tabcolsep}{6pt} 
\renewcommand{\arraystretch}{1} 

\begin{table}[]
\caption{\changed{
Here we evaluate our model in terms of the quality of its estimated illumination environments, compared to previous works and to our model without confidence learning. Our model outperforms prior lighting estimation algorithms, and we see that removing confidence learning significantly degrades performance, thereby demonstrating its value.}
\vspace*{-.1in}
\label{table:light_estimate}}
\begin{center}
\begin{tabular}{ l || c }
Algorithm & $\scalemse$ \\
\hline
\cite{BarronTPAMI2015} & 1.3972   \\
\cite{sfsnetSengupta18} & 1.3252   \\
\hline
Our model &  \cellcolor{Red}    0.6633  \\
Our model w/o confidence learning &  \cellcolor{Orange}    0.8231  \\
\end{tabular}
\end{center}
\vspace*{-.1in}
\end{table}

\subsection{Estimated Lighting}\label{sec:eval-light}
\changed{We also evaluate our model's ability to predict the lighting from a single portrait. Due to the ambiguity between light source strength and surface albedo~\cite{Belhumeur1999}, we use scale-invariant RMSE as our evaluation metric, weighted by the solid angle:}
\begin{equation}
\scalemse(\lightpred, \light) = \min_\alpha \normtwo{ \solidangle\odot(\alpha\targetimagepred - \targetimage) }.
\end{equation}
Where $\solidangle$ contains the solid angle of each pixel in the environment illumination.

\changed{In Table~\ref{table:light_estimate} we evaluate the predicted illuminations produced by our model against two other lighting estimation algorithms~\cite{BarronTPAMI2015,sfsnetSengupta18}, as well as an ablation of our model without confidence learning.
Our method outperforms our ablation and all prior works by a large amount.
See Figure~\ref{fig:light_estimate} for a visualization of these predicted illuminations.
The poor performance of the intrinsic decomposition methods~\cite{BarronTPAMI2015,sfsnetSengupta18} is likely due to their use of spherical harmonics to model illumination, which constrains their output to model only low frequency effects.
Confidence learning's positive contributes is likely because it allows the network to reason about only a part of the lighting from an image patch, thus it helps reducing the speckling in the predicted light behind human subject.}

\section{Real-world Results}\label{sec:real}

\setlength{\tabcolsep}{0pt} 

\newcommand{\wildwidth}{0.14\textwidth} 
\newcommand{\wildwidthd}{0.143\textwidth} 

\begin{figure}[!ht]
\centering
\begin{tabular}{ccc}
Input Image& \multicolumn{2}{c}{Relit Image} \\
\includegraphics[width=\wildwidthd]{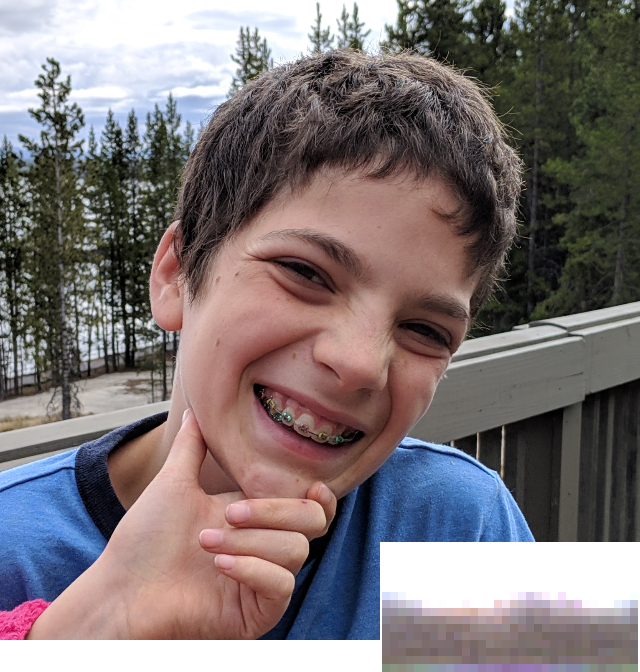}&
\includegraphics[width=\wildwidthd]{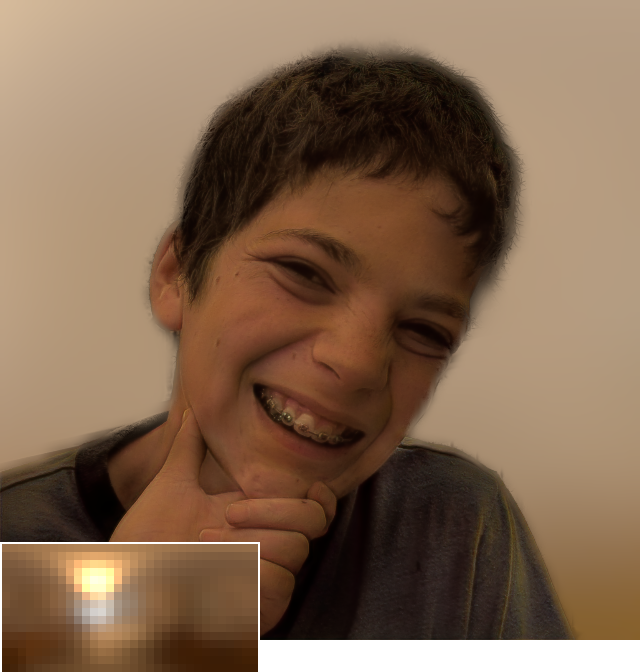}&
\includegraphics[width=\wildwidthd]{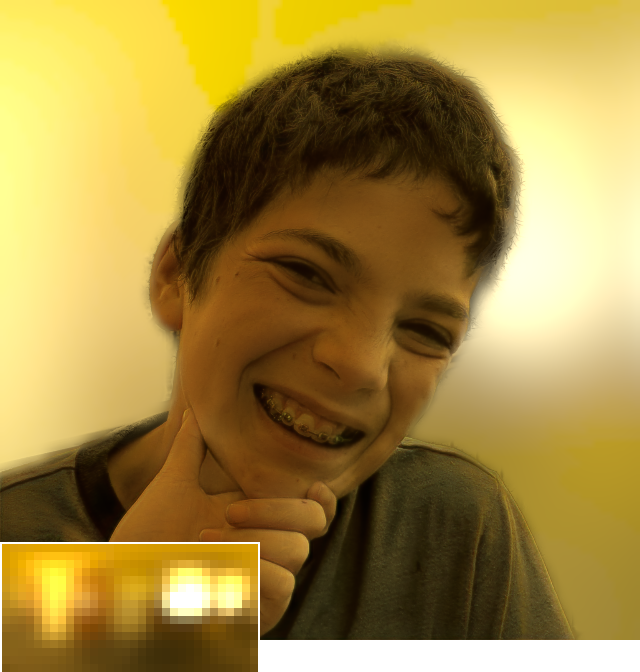}\\
\includegraphics[width=\wildwidthd]{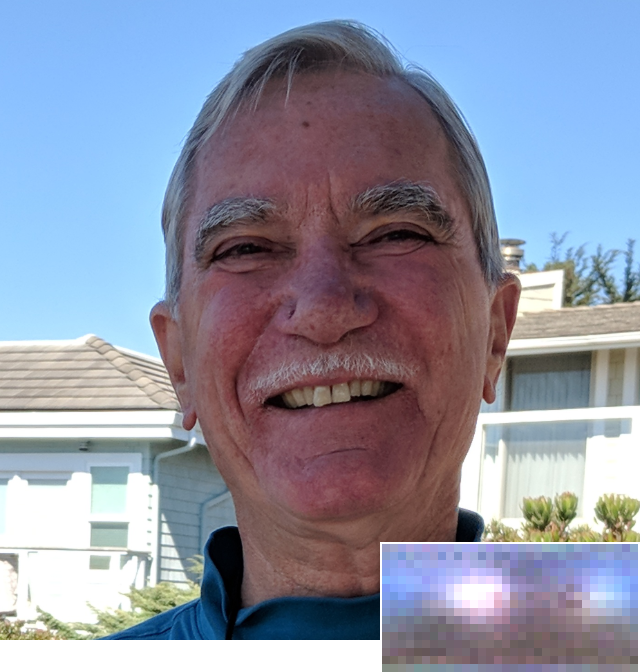}&
\includegraphics[width=\wildwidthd]{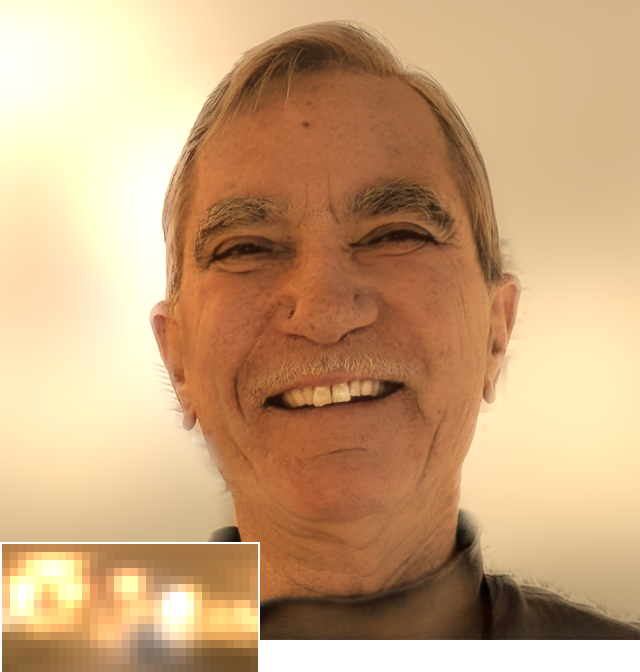}&
\includegraphics[width=\wildwidthd]{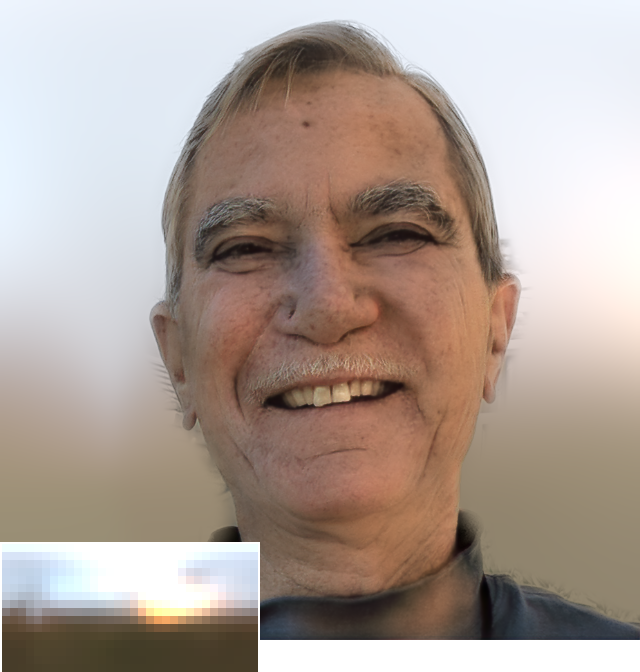}\\
\includegraphics[width=\wildwidthd]{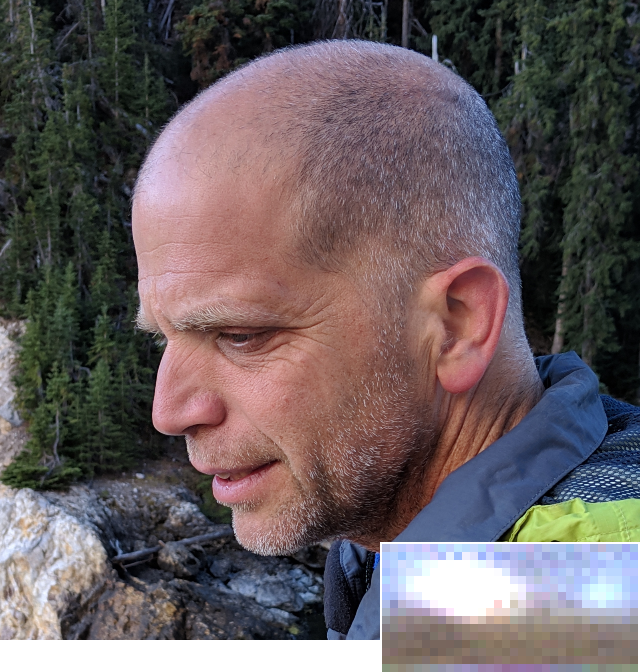}&
\includegraphics[width=\wildwidthd]{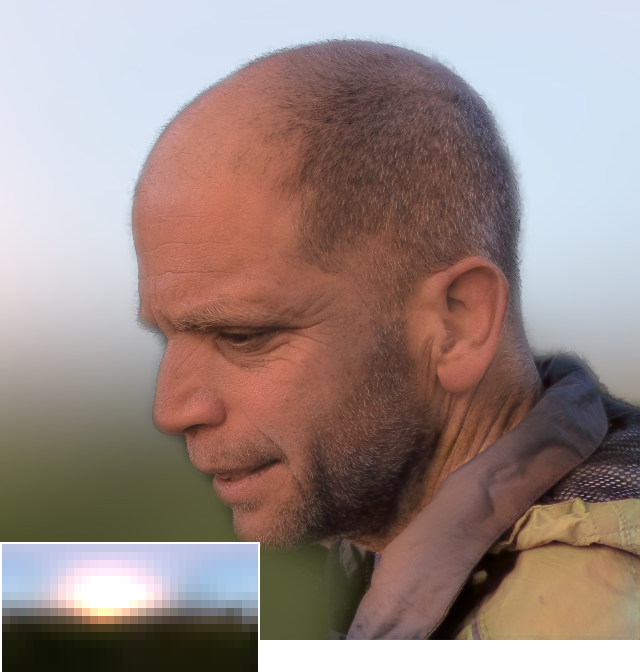}&
\includegraphics[width=\wildwidthd]{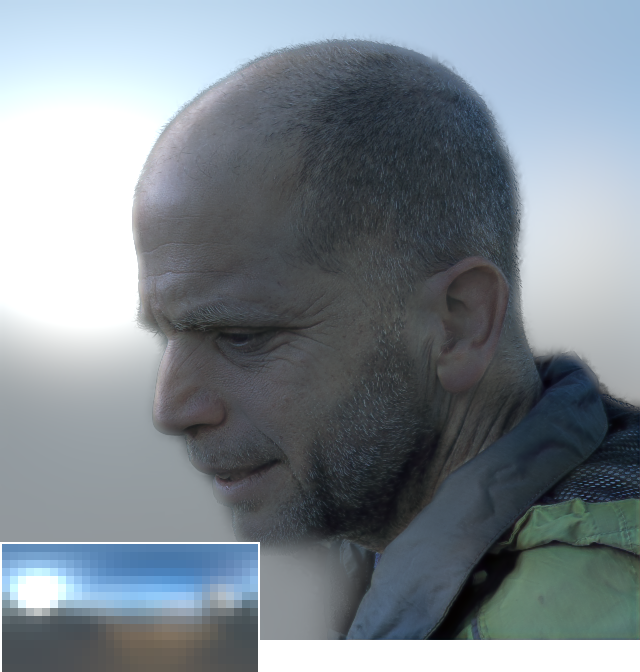}\\
\includegraphics[width=\wildwidthd]{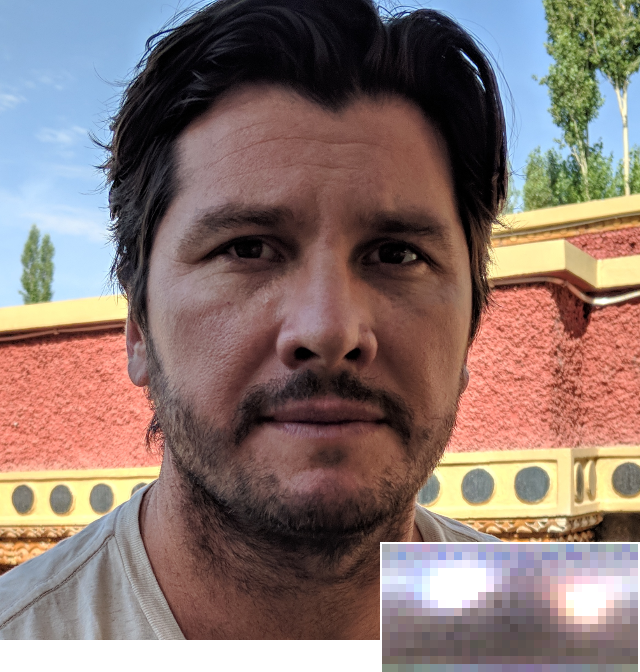}&
\includegraphics[width=\wildwidthd]{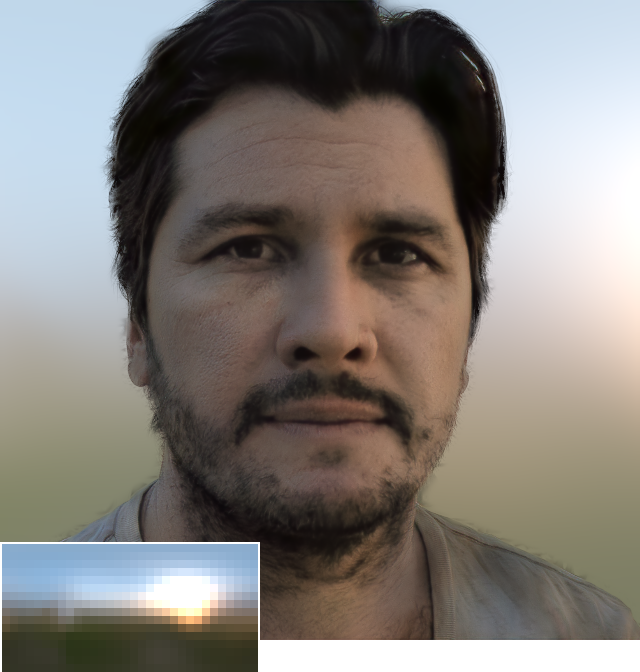}&
\includegraphics[width=\wildwidthd]{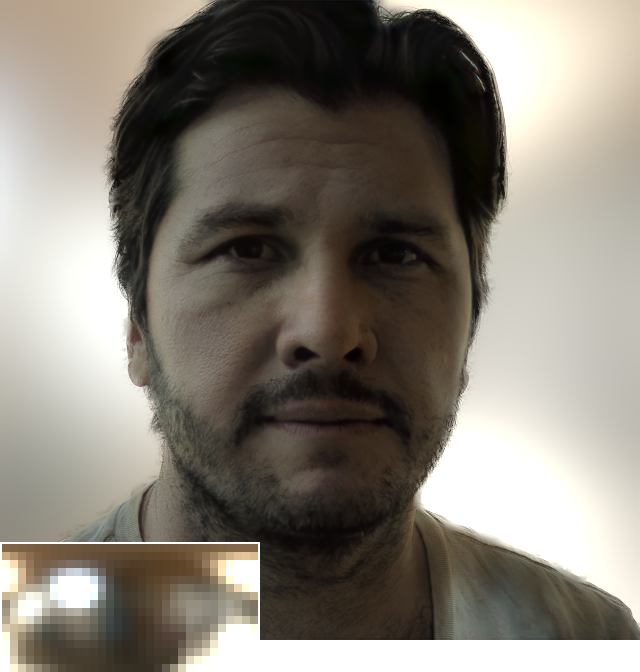}\\
\includegraphics[width=\wildwidthd]{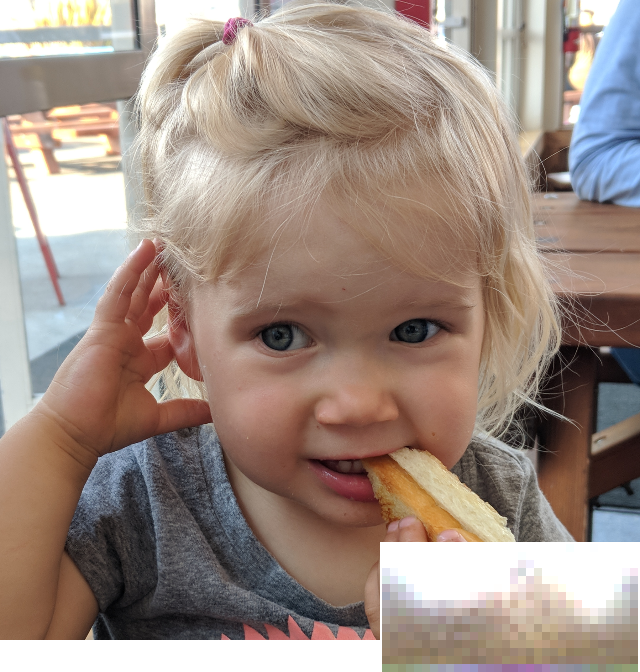}&
\includegraphics[width=\wildwidthd]{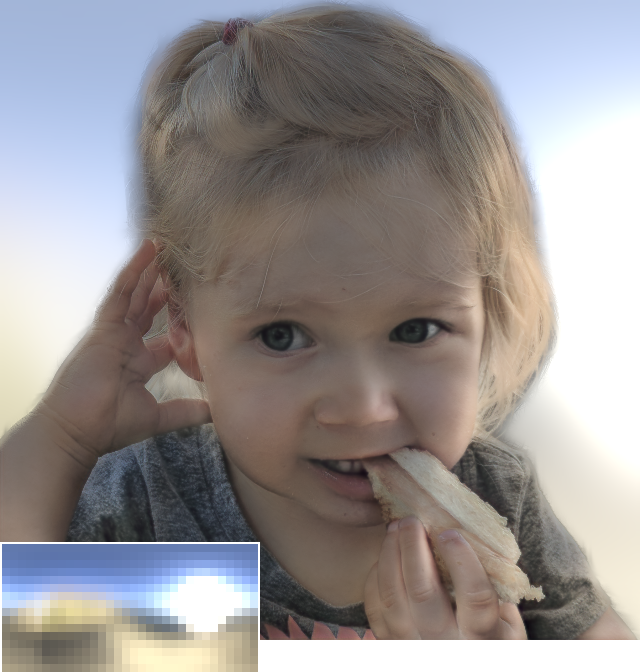}&
\includegraphics[width=\wildwidthd]{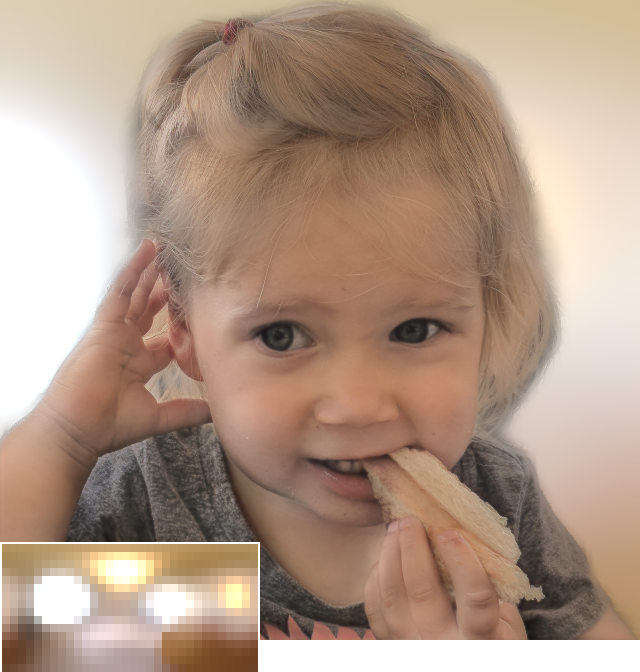}\\
\includegraphics[width=\wildwidthd]{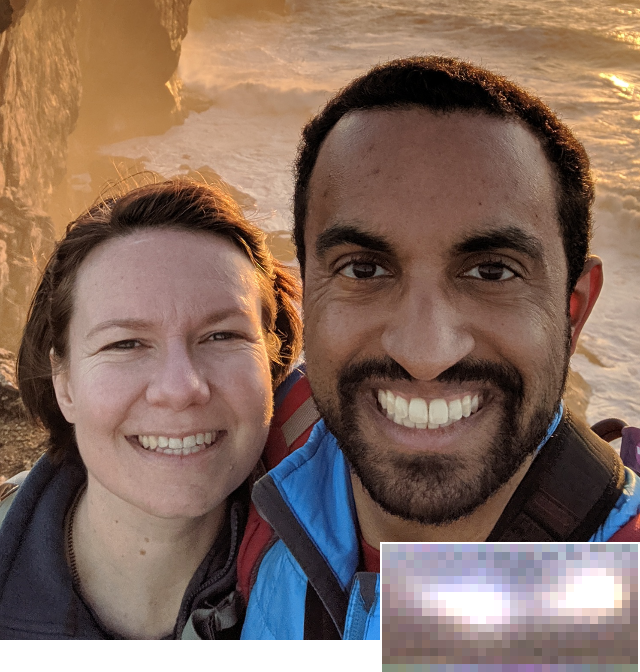}&
\includegraphics[width=\wildwidthd]{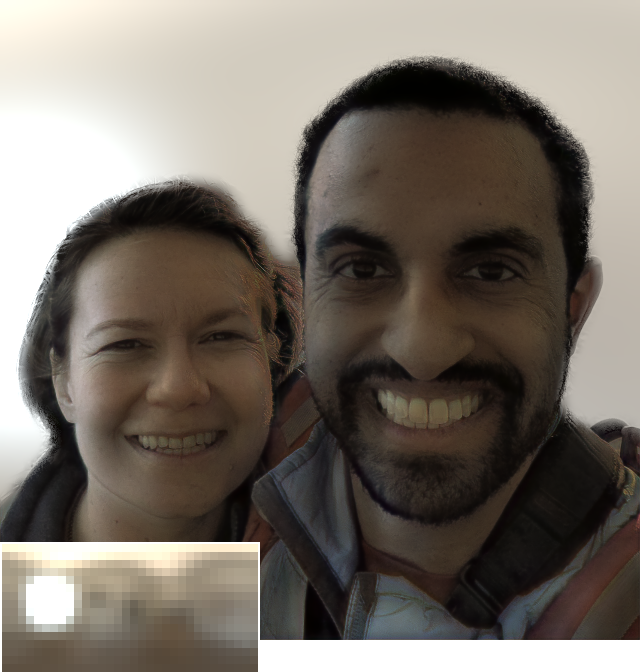}&
\includegraphics[width=\wildwidthd]{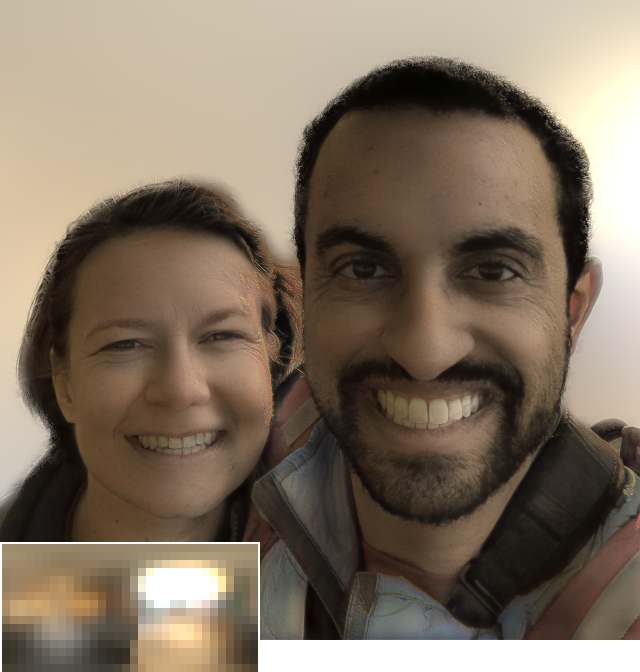}\\
\end{tabular}
\vspace*{-.1in}
\caption{
Examples of our model being used for complete relighting on ``in the wild'' portrait images taken from the Portrait-mode and HDR+ dataset~\cite{Wadhwa2018,Hasinoff2016}.
Using only an RGB input image (column 1) and the portrait mask produced by \cite{Wadhwa2018}, we are capable of producing relit images according to
arbitrary environment illuminations (columns 2 and 3) for a wide variety of subjects and viewing angles.
}
\label{fig:real-complete}
\vspace*{-.2in}
\end{figure}

\begin{figure}[!ht]
  \centering
\begin{tabular}{ccc}
Input Image& \multicolumn{2}{c}{Relit Image} \\
\includegraphics[width=\wildwidth]{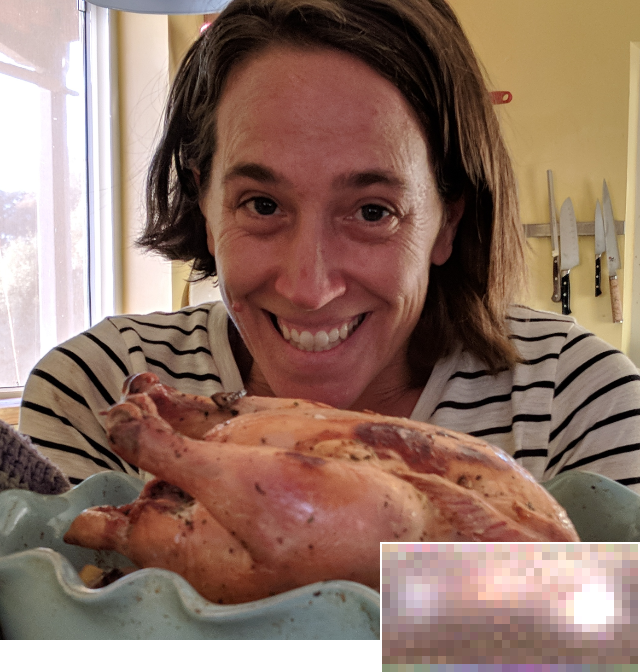}&
\includegraphics[width=\wildwidth]{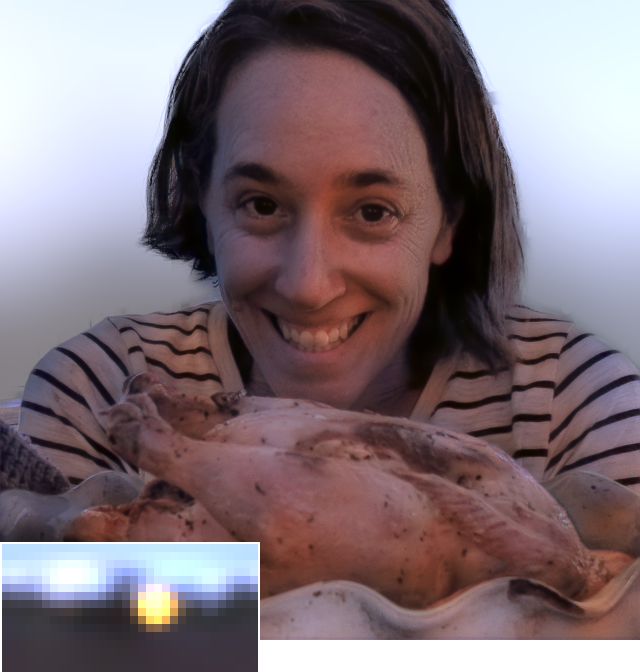}&
\includegraphics[width=\wildwidth]{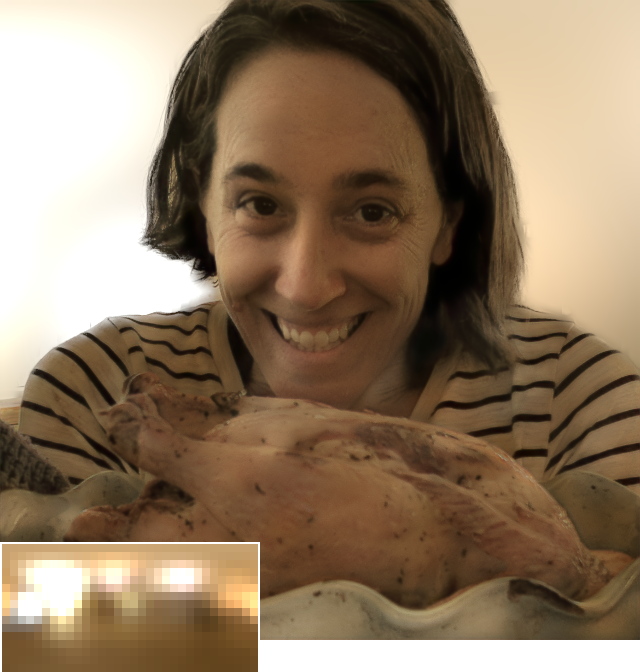}\\
\includegraphics[width=\wildwidth]{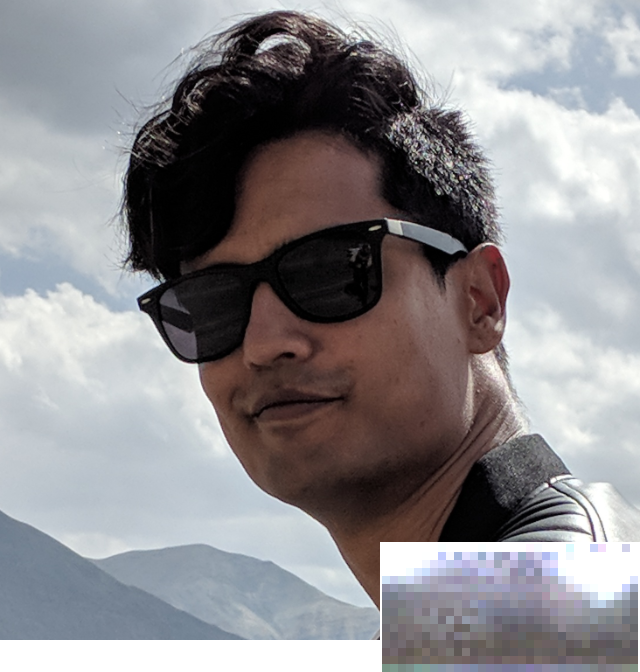}&
\includegraphics[width=\wildwidth]{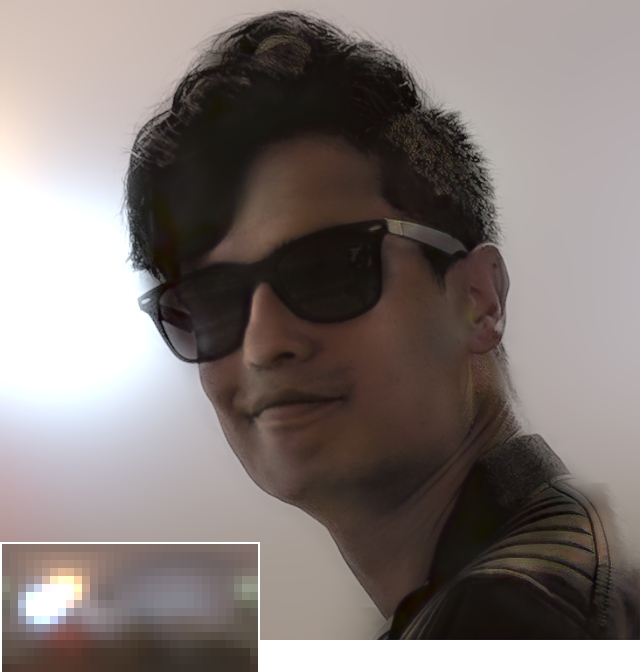}&
\includegraphics[width=\wildwidth]{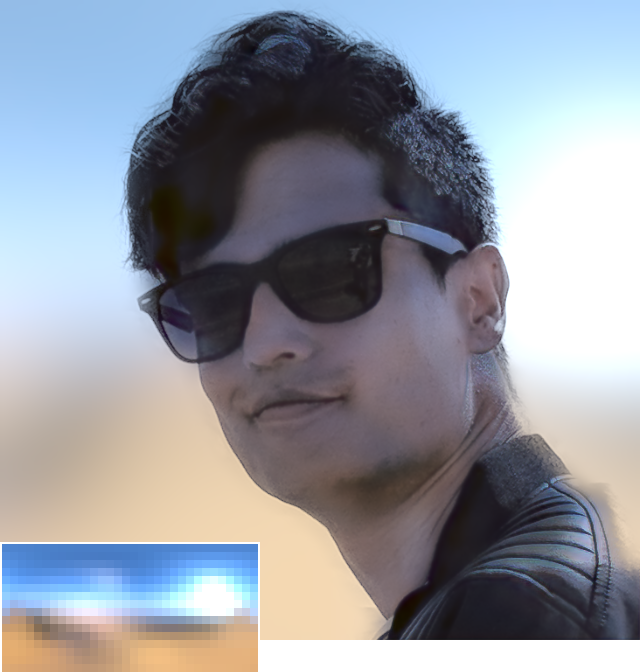}
\end{tabular}\vspace*{-.1in}
\caption{
Portrait images often contain subjects holding objects or wearing accessories such as sunglasses.
Despite not having seen such objects and accessories during training, our model generalizes reasonably to these images.
}\vspace*{-.1in}
\label{fig:real-unknown}
\end{figure}

\begin{figure}[!ht]
  \centering
\begin{tabular}{ccc}
Input Image& \multicolumn{2}{c}{Relit Image} \\
\includegraphics[width=\wildwidth]{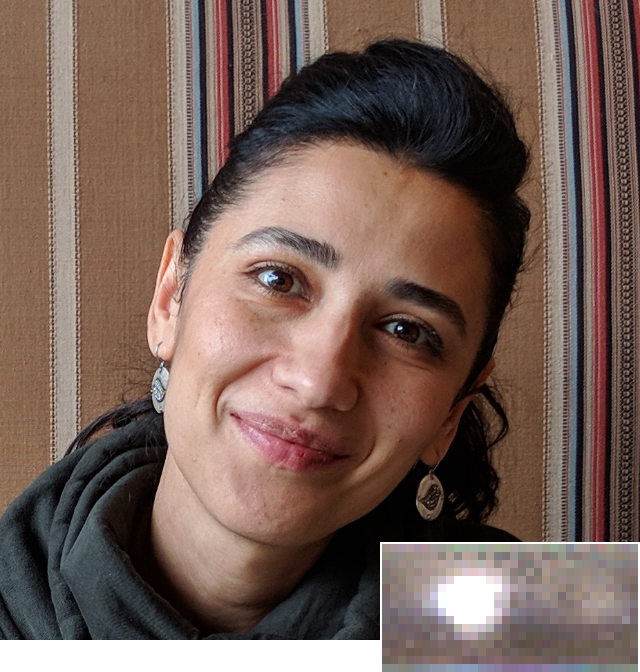}&
\includegraphics[width=\wildwidth]{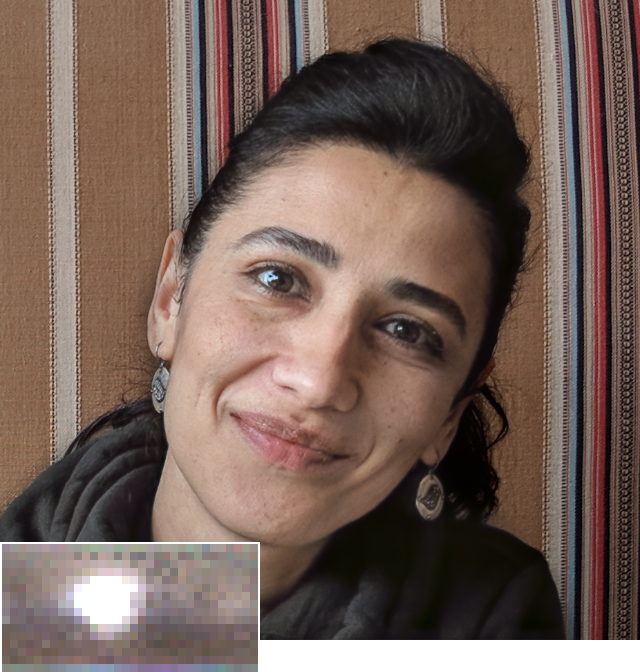}&
\includegraphics[width=\wildwidth]{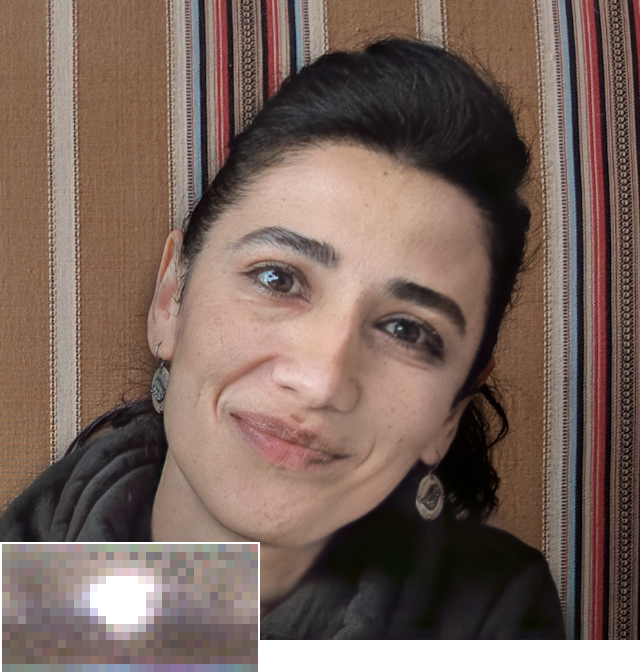}\\
\includegraphics[width=\wildwidth]{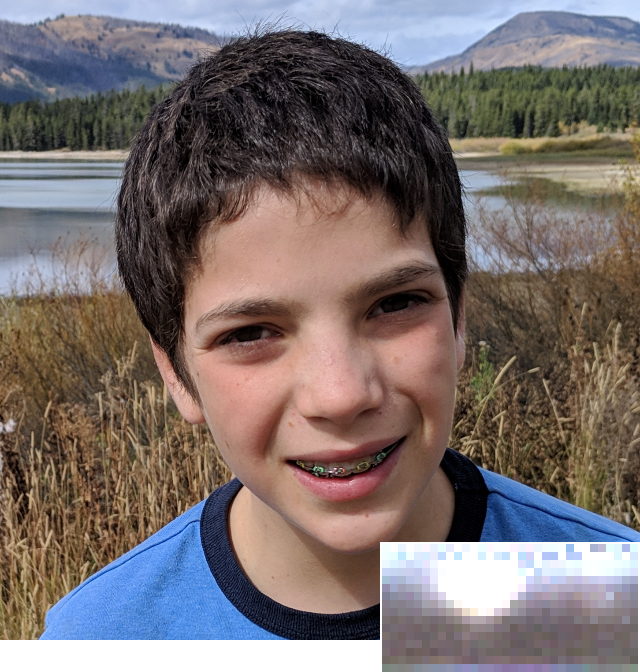}&
\includegraphics[width=\wildwidth]{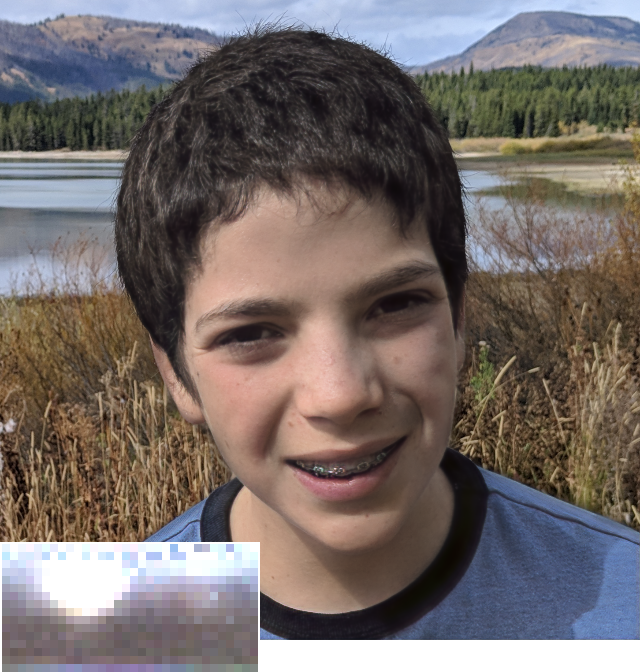}&
\includegraphics[width=\wildwidth]{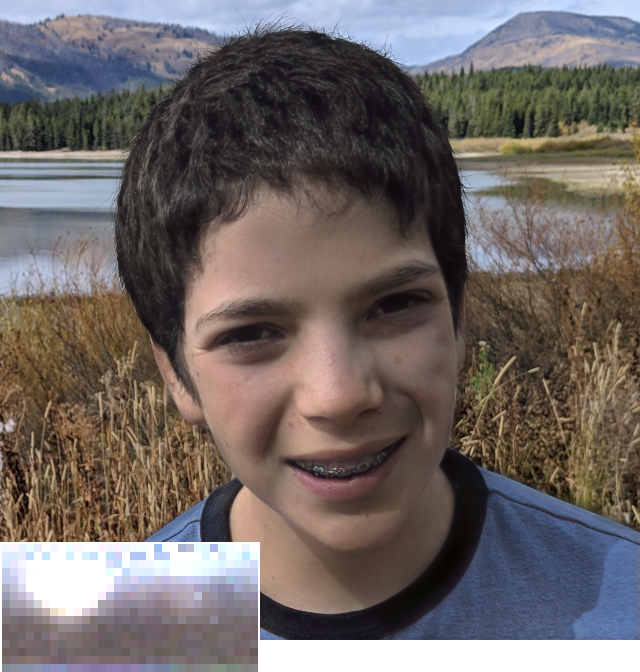}
\end{tabular}\vspace*{-.1in}
\caption{
Here we present instances of illumination retargeting on ``in the wild'' portrait images.
The illumination of the input image is recovered by our model, manually rotated by the user, and then this retargeted illumination is used to re-render the image.
}\vspace*{-.1in}
\label{fig:real-retarget}
\end{figure}

\begin{figure}[!ht]
  \centering
\begin{tabular}{ccc}
Reference Image & Input Image& Relit image \\
\includegraphics[width=\wildwidth]{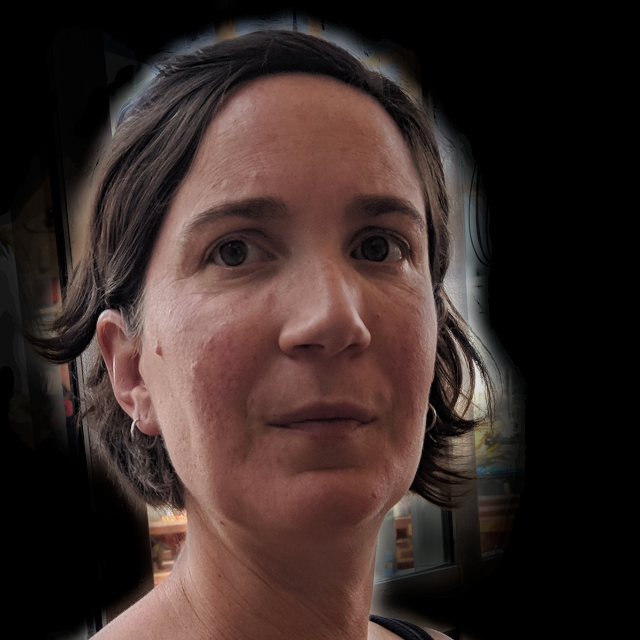}&
\includegraphics[width=\wildwidth]{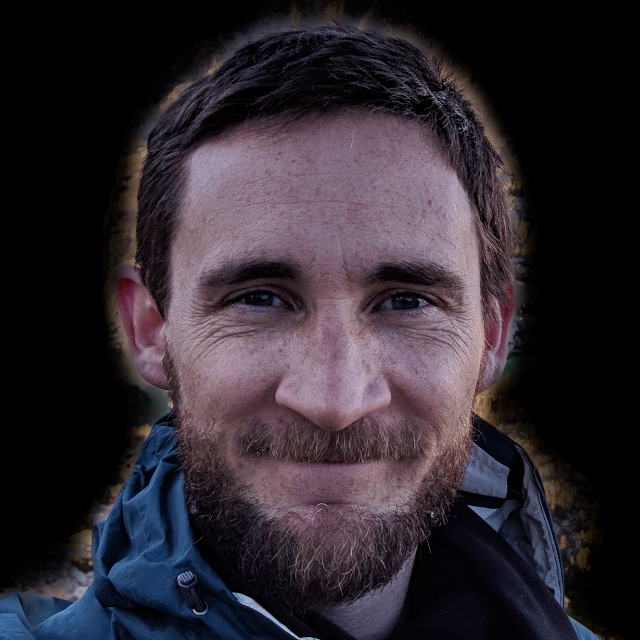}&
\includegraphics[width=\wildwidth]{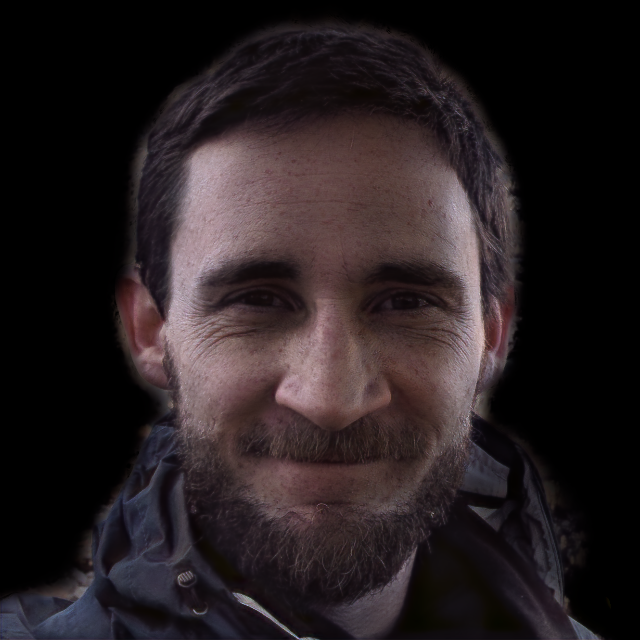}\\
\includegraphics[width=\wildwidth]{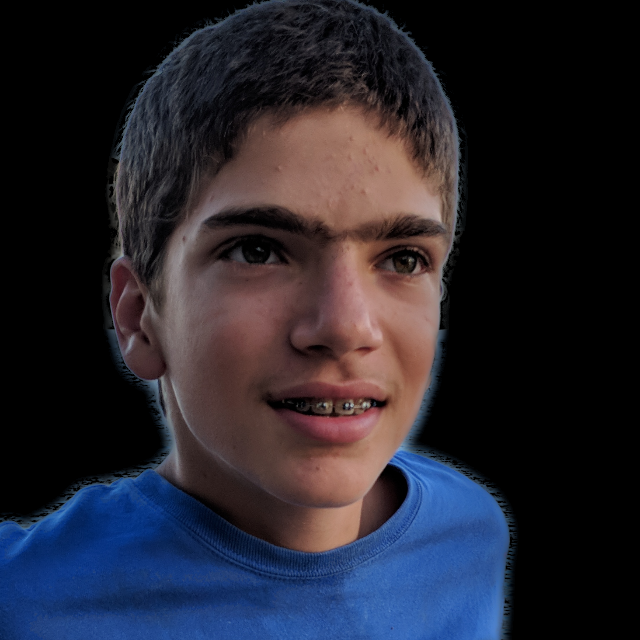}&
\includegraphics[width=\wildwidth]{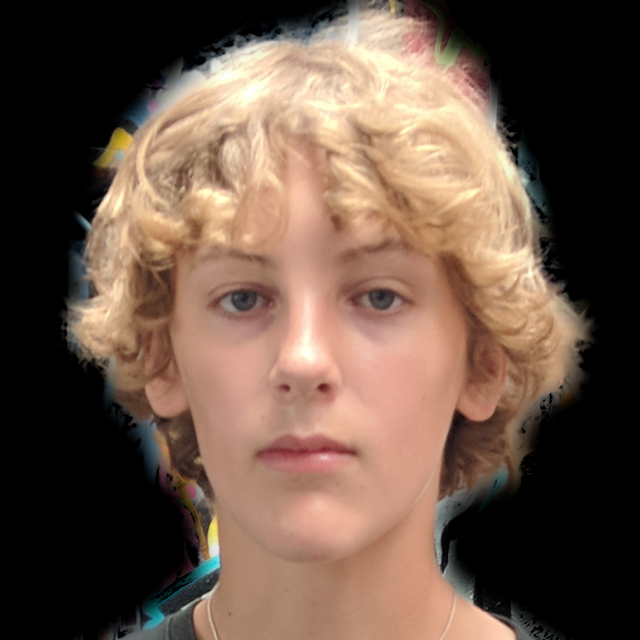}&
\includegraphics[width=\wildwidth]{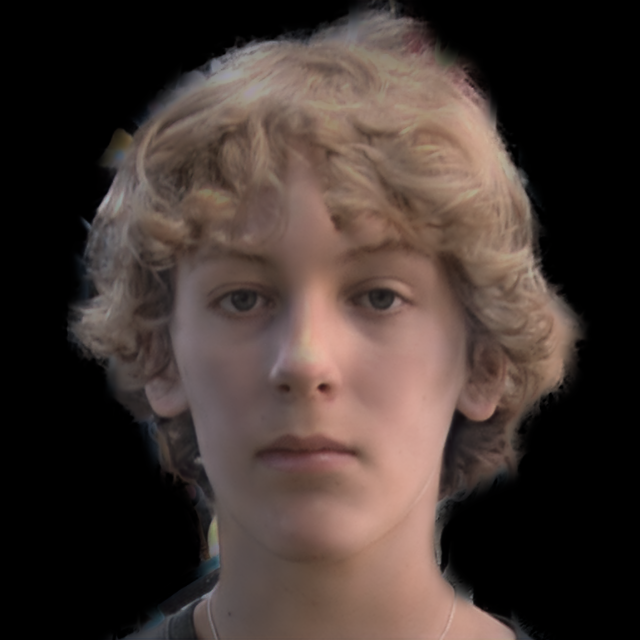}
\end{tabular}\vspace*{-.1in}
\caption{
\changed{Our model can be used for lighting transfer, by using the predicted light of a reference image to relight some input image.}
}\vspace*{-.1in}
\label{fig:real-transfer}
\end{figure}

We additionally evaluate our model on a wide variety of portraits image taken with a standard cell phone, using the Portrait-mode and HDR+ dataset~\cite{Wadhwa2018,Hasinoff2016}.
Despite the small number of subjects in our training data, our model generalizes well to these varied images taken ``in the wild''.

{\bf Complete relighting: } Figures~\ref{fig:teaser} and~\ref{fig:real-complete} show the results of using our model for the task of complete relighting --- replacing the illumination of the scene.
The foreground masks produced by Wadhwa \etc~\shortcite{Wadhwa2018} are used to isolate the portrait's subject before it is used as input to our model, and our renderings are produced by compositing the relit foreground over a synthetic rendering of the target illumination.
Despite significant variation in the skin tones and ages of the subjects, and in the camera angle and initial illumination, our model is capable of producing realistically relit renderings.
Our model is able to process input images that exhibit complex appearance effects such as specularities, soft-shadows and sub-surface scattering, and successfully reproduces those effects in its output renderings.
While our network is trained with only viewing angles within a $20^\circ$ cone around the frontal direction, our method appears to generalize to images with more slanted viewing directions (e.g. the second row of \fig{fig:real-complete}).
Our method also generalizes in terms of the age of the subject, as though our training data consists only of people aged 20 through 40, we are able to process images of children and seniors (e.g. the last two rows of \fig{fig:real-complete}).
The supplementary material shows results on about 60 additional examples, demonstrating the generality of our model.  

In Figures~\ref{fig:teaser} and~\ref{fig:real-unknown}, we see that our network is capable of handling portraits in which the subject contains objects that are not present in the training set (food, hats, and sunglasses).
The network appears to relight and recolor these objects in a way similar to skin or clothing, which results in plausible looking renderings.

\begin{figure}[!ht]
  \centering
\begin{tabular}{ccc}
Input Image& \multicolumn{2}{c}{Relit Image} \\
\includegraphics[width=\wildwidth]{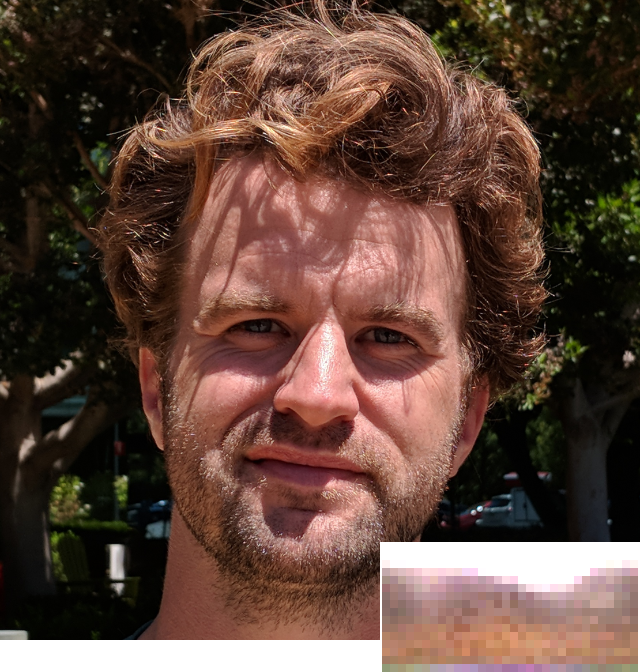}&
\includegraphics[width=\wildwidth]{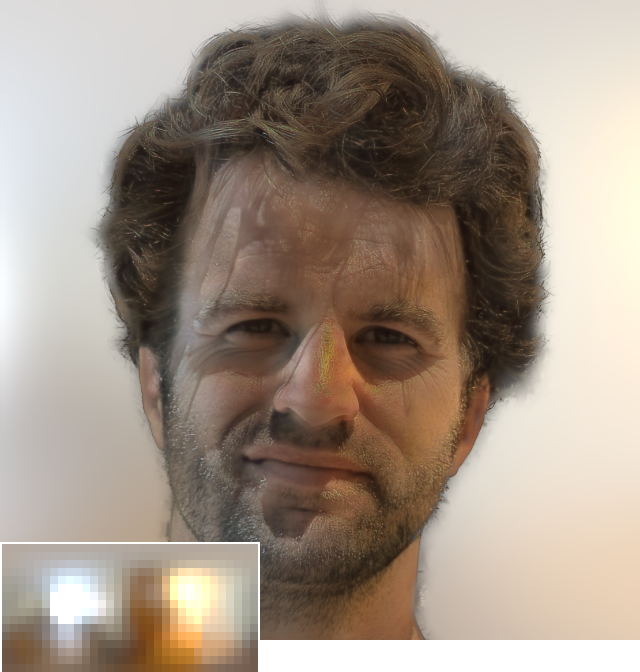}&
\includegraphics[width=\wildwidth]{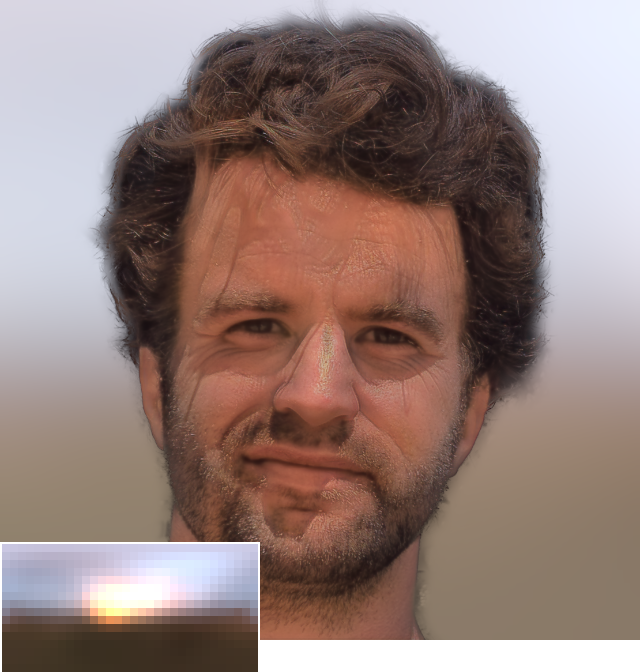}\\
\includegraphics[width=\wildwidth]{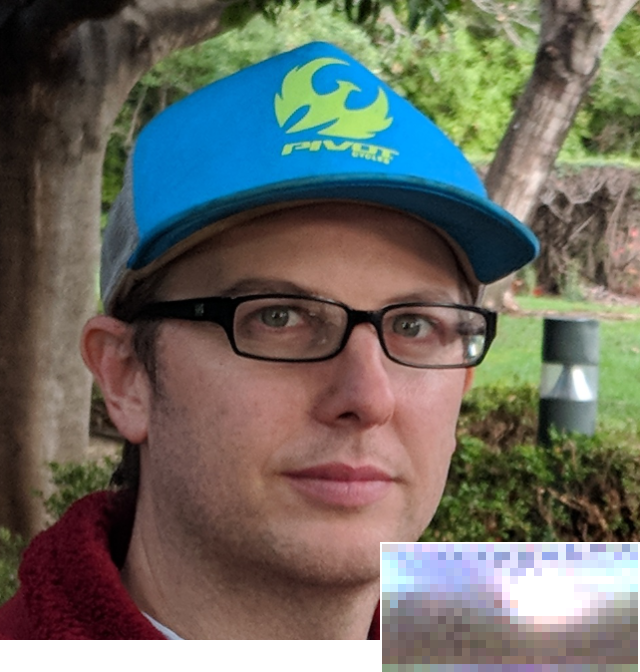}&
\includegraphics[width=\wildwidth]{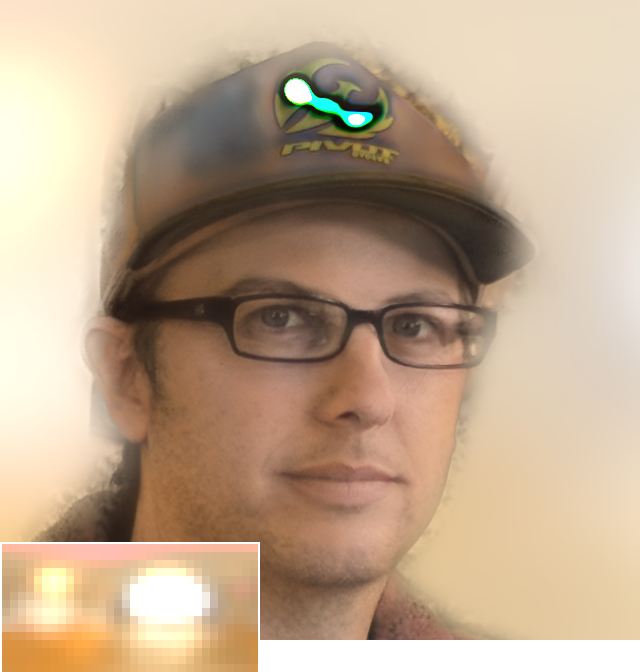}&
\includegraphics[width=\wildwidth]{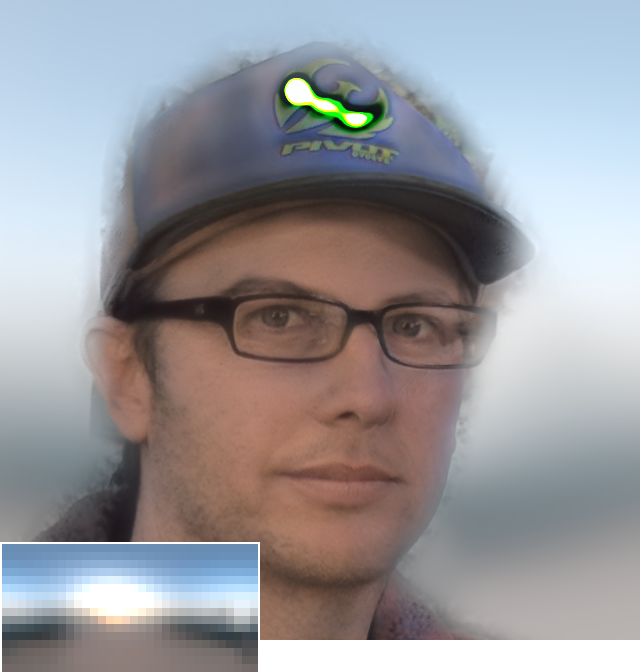}
\end{tabular}\vspace*{-.1in}
\caption{
When presented with images which deviate significantly from our training data, our model may begin to generalize poorly.
Input images lit under harsh illuminations that result in hard shadows, specularities, or saturated pixels may result in artifacts in the output renderings (top).
Additionally, \changed{subjects wearing accessories with highly saturated colors may result in artifacts (bottom).}
}
\label{fig:real-limit}
\vspace*{-.1in}
\end{figure}

{\bf Illumination retargeting: } Because our model produces an estimation of the illumination of the input image, and because of our use of self-supervision during training, our model can be used to ``retarget'' the input illumination and render new portraits in which the original lighting is rotated, as shown in \fig{fig:real-retarget}.
When rendering these examples, instead of replacing the background of the image with the environment map, we composite our relit image back onto the background of the input image using the same mask used to extract the foreground subject.

{\bf Light transfer: } \changed{Though our model was not designed for the task of transferring the lighting from one portrait to another, it can be used for lighting transfer: we predict the lighting from one portrait and then relight another image according to that predicted lighting. As we can see in Figure~\ref{fig:real-transfer}, despite different hair colors or facial expression, our algorithm is capable of producing reasonable lighting transfer results.}

{\bf Limitations: } While our method performs well in most cases, it is not without limitations, as can be seen in \fig{fig:real-limit}. Our model generalizes poorly when presented with input images that contain hard shadowing, sharp specularities, or saturated pixels, as these phenomena are underrepresented in our training data. \changed{It might be possible to fix this problem by adding more training images with hard shadows and sharp specularities.}
Additionally, our model struggles when presented with accessories containing saturated colors, which are not present in our training data. \changed{It may be possible to use semantic or geometric understanding to ameliorate this issue.}

\section{Conclusion}
\label{sec:conclude}

We have presented a learning-based technique for single image portrait relighting: taking a single RGB image of a human subject in an unconstrained environment and modifying it to appear as though it were illuminated by a different environment.
To train our model we have captured a light stage database of a small number of individuals imaged under directional light sources which, when combined with abundantly-available natural environment maps, allows us to generate a large amount of realistic training data for relighting.
We have presented a novel neural network architecture for predicting the illumination of the original environment alongside a relit output image in a new target environment, and we have demonstrated the value of self-supervision during training.
Our technique outperforms state-of-the-art relighting techniques on the validation set of our dataset, and performs well in practice on another dataset of hundreds of real-world unconstrained portrait photos taken with an ordinary cellphone camera.
The high-quality renderings produced by our technique, combined with its generality and speed ($160$ milliseconds per $640 \times 640$ image) suggests that our model may enable compelling consumer-facing photographic relighting applications.
Our model may also enable other computer vision applications, perhaps by being used to synthesize additional training data for tasks such as facial recognition of 3D reconstruction.
We expect that additional training data will enable robustness to some of our model's current limitations, such as very hard shadows and sharp specularities.

\begin{acks}
This work was funded in part by a Jacobs Fellowship, 
the Ronald L. Graham Chair, 
and the UC San Diego Center for Visual Computing.
Thanks to Zhixin Shu and Yichang Shih for the help on baseline algorithms,
to Jean-Fran\c{c}ois Lalonde for providing the indoor lighting dataset,
to Peter Denny for coordinating dataset capture,
and to all the anonymous volunteers in the dataset.
\end{acks}

\bibliographystyle{ACM-Reference-Format}
\bibliography{ref}
\end{document}